\documentclass{IEEEtran}
\IEEEoverridecommandlockouts
\usepackage{cite}
\usepackage{amsmath,amssymb,amsfonts}
\usepackage{algorithmic}
\usepackage{graphicx}
\usepackage{textcomp}
\usepackage{xcolor}
\usepackage{setspace}
\def\BibTeX{{\rm B\kern-.05em{\sc i\kern-.025em b}\kern-.08em
		T\kern-.1667em\lower.7ex\hbox{E}\kern-.125emX}}
\usepackage{subfigure}
\newtheorem{my_theorem}{Theorem}

\newtheorem{my_lemma}{Lemma}

\newtheorem{my_proposition}{Proposition}

\newcommand*{\J}{\jmath}%
\usepackage{enumitem}
\usepackage{array}

\begin{document}
	\title{Exact Sum Distribution of $\alpha$-$\eta$-$\kappa$-$\mu$ Fading Channels for Statistical Performance Analysis of RRS-Based Wireless Transmission}
	\author{Rayees Ahmad Wani,~\IEEEmembership{Graduate Student Member,~IEEE}, Rajeev Rajagopal,  Pranay Bhardwaj,~\IEEEmembership{Graduate Student Member,~IEEE},  El Mehdi Amhoud,~\IEEEmembership{Senior Member,~IEEE}, and     S.~M.~Zafaruddin,~\IEEEmembership{Senior Member,~IEEE} 
		\thanks{A conference version of the paper, presenting the analysis on the exact representation of the \(\alpha\)-\(\eta\)-\(\kappa\)-\(\mu\) fading model for a single link, was published in the 2023 IEEE Globecom Workshops (GC Workshops), Kuala Lumpur, Malaysia, December 2023 \cite{Bhardwaj2023gc}.}
		\thanks{ Rayees Ahmad Wani (p20240042@pilani.bits-pilani.ac.in), Rajeev Rajagopal (f20201237@pilani.bits-pilani.ac.in),  and S.~M.~Zafaruddin (syed.zafaruddin@pilani.bits-pilani.ac.in)  are with the Department of Electrical and Electronics Engineering, Birla Institute of Technology and Science, Pilani, Pilani-333031, Rajasthan, India.} \thanks{Pranay Bhardwaj (pranaybhardwaj@nyu.edu) is with Engineering department, New York University Abu Dhabi.} \thanks{El Mehdi Amhoud (ElMehdi.Amhoud@um6p.ma) is with Univeristy Mohammed VI Polytechnic, Morocco.}
		\thanks{This work was supported in part by the Anusandhan National Research Foundation (ANRF), Department of Science and Technology (DST), Government of India under  Core Research Grant CRG/2023/008040 and MATRICS Grant MTR/2021/000890, and in part by the Ministry of Education, Government of India Scheme for Promotion of Academic and Research Collaboration (SPARC Phase III) for the project  P3183 “Multiagent AI for Controlled Sensing and Communication”, and in part by the BITS Pilani SPARKLE project. }
	}
	
	\maketitle
	
	\thispagestyle{empty}

	\maketitle 

	\begin{abstract}
Reconfigurable refractive surface (RRS) is an efficient alternative to holographic multiple input multiple output (HMIMO) systems that can serve as a transmission unit operating in the signal refraction mode. RRS transmissions experience near-field propagation due to the proximity of the transmission feed and far-field propagation for the user located farther from the transmitting unit. There is limited research on the effect of channel fading on far-field users in RRS-based transmissions. In this paper, we conduct an exact statistical analysis of RRS-based transmission considering  $\alpha$-$\eta$-$\kappa$-$\mu$ fading model for the far-field user and the near-field effect of transmission feed. First, we show that the exact statistical analysis for the RRS transmission over $\alpha$-$\eta$-$\kappa$-$\mu$ fading model consists of multiple infinite-series representations with multivariate Fox-H function. Next, we develop a novel approach to derive the density and distribution functions for the resultant fading channel of the  RRS system in terms of multivariate Fox-H functions without involving infinite series approximations for tractable performance analysis. We present the exact outage probability and average bit-error-rate (BER) performance of single-element and multiple-element RRS systems to validate the proposed analysis further. We also obtain the diversity order of the system by analyzing the outage probability at a high signal-to-noise ratio (SNR). Computer simulations are used to demonstrate the relevance of the developed statistical results for RRS-based wireless systems over  the generalized fading  model for a comprehensive performance evaluation.
	\end{abstract}			

	\section{Introduction}
	Holographic Multiple Input Multiple Output (HMIMO) technology is rapidly becoming a key element in future wireless networks, leveraging a dense array of small antennas or reconfigurable elements within a compact space to form a spatially continuous aperture \cite{Huang2020},\cite{Deng2021}. However, implementing HMIMO using traditional phased arrays poses significant challenges, particularly due to the requirement for hundreds of high-resolution phase shifters, which leads to high power consumption. Meta-surfaces offer a promising solution for 6G wireless networks by enabling reconfigurable signal propagation. The reconfigurable intelligent surface (RIS) is generally positioned between a source and destination to act as a reflective unit, creating a more favorable propagation environment. Recently, the meta-surface, used as a transmission unit near the feed in a signal refraction mode, has emerged as an efficient alternative to conventional HMIMO systems. A reconfigurable refractive surface (RRS) is an ultra-thin surface embedded with sub-wavelength elements \cite{Yu2022},\cite{10000596}, that can refract incident electromagnetic waves. The RRS uses reconfigurable phase shifts, which are tuned by controlling the biased voltages on the diodes to achieve desired beamforming. 
	
	Although the use of RIS has been extensively studied in the literature, the feasibility of RRS for wireless communication is in the nascent stage \cite{Zeng2022,Zeng2024, Li2024}. The authors in \cite{Zeng2022} analyzed the data rate and power consumption  of a RRS-aided downlink transmission for a single user with a single base station (BS) functioning as an RRS. They demonstrated that the RRS consumes less power than the phased array when the UE is in the far-field region of the BS, establishing the RRS as an energy-efficient solution for holographic MIMO. In \cite{Zeng2024}, the authors extended the concept to RRS-enabled multiuser HMIMO system in which a single BS employs an RRS as the transmit antenna to serve multiple users. They analyzed the system capacity and formulated an optimization problem to enhance the energy efficiency of the RRS-aided system by optimizing the number of RRS elements.

	The sixth generation (6G) and beyond wireless communication systems are expected   to utilize conventional radio frequency (RF) and high-frequency spectrum bands such as millimeter-wave (mmWave), and terahertz (THz) \cite{Dang_2020_nature,Akyildiz_2020}.  Initial research proposed classical Rayleigh and Rice distributions for small-scale fading amplitudes for the mmWave band, specifically at a carrier frequency of $28$ \mbox{GHz} and $60$ \mbox{GHz} \cite{Thomas1994,Smulders2009,Moon2005,Samimi2016}. However, single-parameter models such as Rayleigh, Rice, Hoyt, and Nakagami-m may not provide enough flexibility to accurately  fit the measurement data in some  propagation scenarios, requiring more generalized and flexible models. 	In the seminal work \cite{Yacoub_2016_alpha_eta_kappa_mu}, M. D. Yacoub proposed the $\alpha$-$\eta$-$\kappa$-$\mu$ model as the most flexible and comprehensive wireless channel fading model. Later, Silva \emph{et al.} \cite{Silva_2020_alpha_eta_kappa_mu} improved the mathematical representation of the PDF and CDF of the channel envelope.  The authors in \cite{Marins2019_alpha_eta_kapp_mu} examined the applicability of  Rayleigh, Rice, $\alpha$-$\mu$, $\kappa$-$\mu$, $\eta$-$\mu$,  and $\alpha$-$\eta$-$\kappa$-$\mu$  for mmWave transmission in the range from $55$ \mbox{GHz} to $65$ \mbox {GHz} through extensive measurement campaign in an indoor environment. They found that $\alpha$-$\mu$, $\kappa$-$\mu$, and $\eta$-$\mu$ can be applied to most situations. However, these models fail to provide a good fit in some of the intricate occasions even in the lower frequency range ($55$ \mbox{GHz} to $65$ \mbox {GHz}) and thus expected to fail at a much higher mmWave/THz frequency in most of the propagation scenarios. Further, \cite{Marins2019_alpha_eta_kapp_mu} demonstrated that the $\alpha$-$\eta$-$\kappa$-$\mu$  is the best option fitting over a wide range of situations, including intricate and rare events.There are research works using the  $\alpha$-$\eta$-$\kappa$-$\mu$ model for wireless communications \cite{Li_2017_alpha_eta_kappa_mu,Mathur_2018_alpha_eta_kappa_mu,Jia2018_alpha_eta_kapp_mu,Moualeu_2019_alpha_eta_kappa_mu,Singh2019_alpha_eta_kapp_mu,Ai_2020_alpha_eta_kappa_mu,Al2020_alpha_eta_kapp_mu,Saraereh2020_alpha_eta_kapp_mu,Soni2020_alpha_eta_kapp_mu,Vishwakarma2021_alpha_eta_kapp_mu,GOSWAMI2019_alpha_eta_kapp_mu,Kavaiya2020_alpha_eta_kapp_mu}. 	Further, 
	the statistical results of the $\alpha$-$\eta$-$\kappa$-$\mu$ model for the single link  presented in \cite{Yacoub_2016_alpha_eta_kappa_mu} \cite{Silva_2020_alpha_eta_kappa_mu} contains an infinite series comprising of mathematical functions such as regularized hypergeometric function and generalized Laguerre polynomial. These functions have  complex mathematical formulations and do not present a straightforward insight into propagation channel behavior.
	  It should be mentioned that most of the existing literature \cite{Li_2017_alpha_eta_kappa_mu,Mathur_2018_alpha_eta_kappa_mu,Jia2018_alpha_eta_kapp_mu,Moualeu_2019_alpha_eta_kappa_mu,Singh2019_alpha_eta_kapp_mu,Ai_2020_alpha_eta_kappa_mu,Al2020_alpha_eta_kapp_mu,Saraereh2020_alpha_eta_kapp_mu,Soni2020_alpha_eta_kapp_mu,Vishwakarma2021_alpha_eta_kapp_mu} presented  analysis of performance metrics such outage probability, average bit-error rate, and ergodic capacity   using Meijer's G-function and/or Fox's H-function with infinite-series presentation.

	The proof of concept for RRS-based transmissions requires further investigation, particularly regarding its performance in fading channels. In \cite{Li2024}, both reflection and refraction characteristics of the meta-surface were examined for two-user transmission under NOMA with Nakagami-m fading. The authors developed an upper bound on the ergodic capacity by using moments of the Nakagami-m distribution to avoid the derivation of  distribution of the sum of i.i.d. random variables. An exact statistical analysis for RRS-based transmission over fading channel is required for a better performance assessment.  Statistical performance analysis for RRS-based transmission over $\alpha$-$\eta$-$\kappa$-$\mu$ fading channel is highly motivated due to the generalization of the fading model. The analysis of RRS-based transmission requires derivation for the sum of the product of the near-field  and the far-field channel coefficients. It should be emphasized that there  is no analysis involving the sum of $\alpha$-$\eta$-$\kappa$-$\mu$ model even for the far-field propagation  required in the analysis for MRC/EGC receivers.

This paper develops an exact statistical analysis for RRS-based wireless transmission over the generalized $\alpha$-$\eta$-$\kappa$-$\mu$ fading model for the far-field user and the near-field effect of transmission feed.. The major contributions of the paper are as follows: 
\begin{itemize} 
	\item We show that the PDF of the resultant channel for an $N$-element RRS-based transmission, using the existing representation of the $\alpha$-$\eta$-$\kappa$-$\mu$ fading model, involves an  $N$-infinite series sum over $2N$-variate Fox-H functions. This complex representation requires computations of infinite series and multivariate Fox-H functions. An exact representation of the sum for i.ni.d. $\alpha$-$\eta$-$\kappa$-$\mu$ variates in terms of standard mathematical functions is infeasible. 
	\item We employ a novel approach to derive an exact statistical representation of the $\alpha$-$\eta$-$\kappa$-$\mu$ fading model by presenting the PDF and CDF of the channel envelope using a single trivariate Fox-H function without any infinite series approximation. Multivariate Fox-H functions are widespread in the research community for analyzing the exact and asymptotic performance of wireless systems over generalized fading models. 
	\item We utilize the statistical results of the single channel to analyze the $N$-element RRS-based transmission by expressing the sum of trivariate Fox-H functions in terms of a $3N$-variate Fox-H function. 
	\item To further validate the proposed analysis, we present the exact outage probability and average bit-error-rate (BER) performance of a single-element and multiple-element RRS system subjected to the $\alpha$-$\eta$-$\kappa$-$\mu$ fading model. We also obtain the diversity order of the system by analyzing the outage probability at a high signal-to-noise ratio (SNR). 
	\item The developed statistical results for RRS-based wireless systems are demonstrated through computer simulations over the generalized fading model for a comprehensive performance evaluation. 
	\end{itemize}
	
	The paper is organized as follows: Section II discusses the system description and channel model for RRS transmission. Section III presents statistical results for RRS-based transmission for infinite-series approximation and exact representation subjected to the $\alpha$-$\eta$-$\kappa$-$\mu$ fading model. Section IV develops an exact analysis of the outage probability and average BER for single-element and multiple-element RRS systems. Section V demonstrates computer simulations for RRS performance over various channel conditions. Finally, Section VI concludes the paper.

	\section{System Description and Channel Models}
	We begin by describing the system model for RRS-assisted transmission, which encompasses both near-field and far-field propagation scenarios. Then, we detail the channel and fading model for conducting statistical performance analysis.
	\subsection{System Description}
\begin{figure}[t]
		\centering
		\includegraphics[width=\columnwidth]{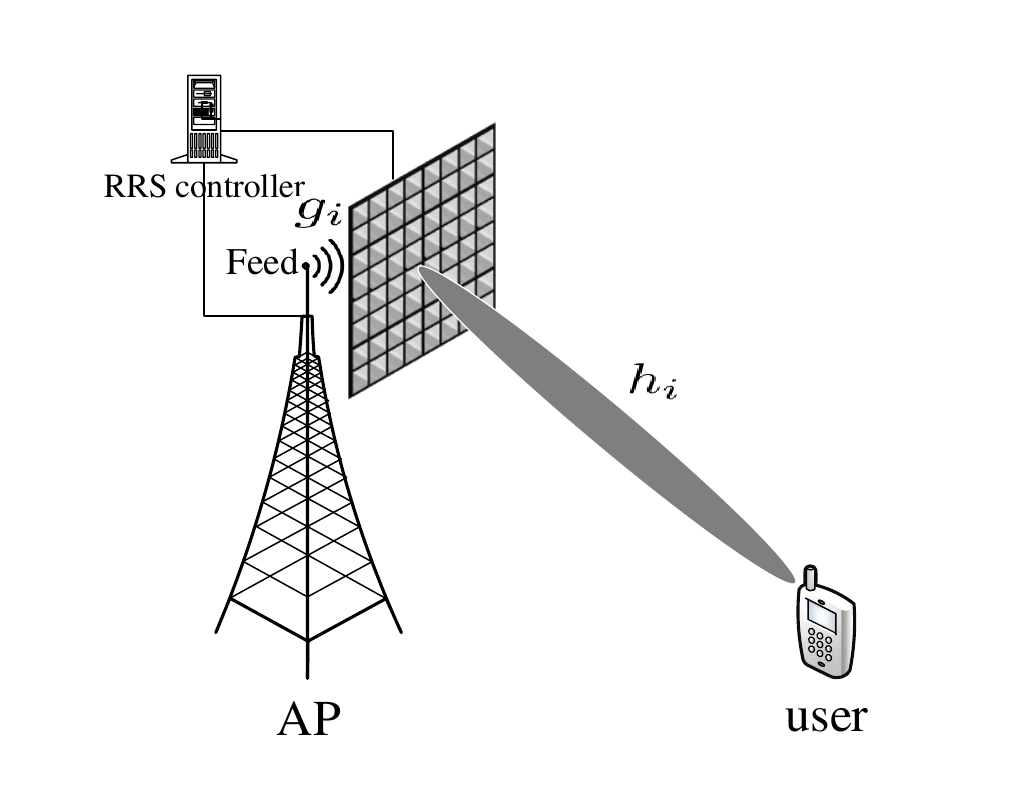} 
		\caption{A schematic diagram  for RRS-based wireless transmission.}
		\label{fig:system_model}
	\end{figure}

	We consider an $N$-element RRS transmission where an access point (AP) communicates with a single user, as shown in  Fig.~\ref{fig:system_model}. An extension to  multi-user transmission can be similarly adopted by applying beam-forming vector at the RRS for multiple signal transmission.  In the single-user configuration, signal $s$ from  the feed is propagated  towards the RRS module within the near-field region. The coordinates of the feed are given as \((0, 0, -d_0)\), while the coordinates of the center point of the $i$-th RRS element are  taken as  $(p_i, q_i, 0)$.  With the near-field propagation,  the channel gain  from the feed to the $i$-th RRS element becomes  $|g_i|=\big[{\int\int_{D_{i}}}\frac{(\alpha+1){d_{o}}^{\alpha+1}}{2\pi}{(d_{o}^2+p^2+q^2)^{\frac{-\alpha+3}{2}}}dpdq\big]^{1/2}$ where  $2(\alpha + 1)$ is the gain of the feed, and the integration region is 
	$D_i = \{(p, q) \in \mathbb{R}^2 : x_i - \frac{\delta x}{2} \leq x \leq x_i + \frac{\delta x}{2}, y_i - \frac{\delta y}{2} \leq y \leq y_i + \frac{\delta y}{2}\}$ \cite{Li2024}.
	Thus, we  express  $g_i$ as the channel coefficient from the feed to the $i$-th RRS as
	\begin{eqnarray}\label{eq:gn1}
	g_{i}=|g_i| e^{j\frac{2\pi}{\lambda}\sqrt{{p_{i}^2+q_{i}^2+d_{0}^2}}}
	\end{eqnarray}
	where  $\lambda$ is the wavelength of prorogation.

	The signal received at a far-end user through $N$-element RRS is given by
	\begin{equation}\label{eq:received equation1}
	y = \sum_{i=1}^{N} \beta_i g_i h_i s +w
	\end{equation}
	where $h_i$ is the far-end  channel coefficient from the $i$-th RRS element to the UE. The channel coefficient 	$h_{i}= h^{l}h_{i}^f$, where $h^l$ is the  path gain   and 	$h_i^{f}=|h_i^f| e^{j \phi_n^f}$ denotes multi path fading with $|h_i^f|$ as the amplitude and $\phi_i^f$ as the phase. The additive white Gaussian noise (AWGN) is modeled  $w\sim {\cal{N}}(0,\sigma_w^2)$.
	Each RRS elements require to compensate the phase so that received signal at the UE is coherently added. We use perfect phase compensator  $\beta_i=\Gamma_ie^{-j\phi_i}$, where $\phi_i=\frac{2\pi}{\lambda} \sqrt{p_i^2 + y_i^2 + d_o^2} + \phi_i^f$ and $\Gamma_i$ is the refraction coefficient of the $i$-th RRS element. Assuming $\Gamma_i=\Gamma$, $\forall i$, we represent \eqref{eq:received equation1} as $y =  h^{l}\Gamma\sum_{i=1}^{N}  |g_i| |h_i^f|s +w$. Hence, the resulting  signal-to-noise ratio  for an $N$-element RRS system can be expressed as:
	\begin{flalign}\label{eq:snr1}
	\gamma = \bar{\gamma} \left( \sum_{i=1}^{N} |g_i| |h_i^f| \right)^2
	\end{flalign}
	where $\bar{\gamma}=\frac{  |h^l|^2 |\Gamma|^2 P}{\sigma_w^2}$ is the SNR without fading.
	As seen from \eqref{eq:snr1}, the statistical analysis for RRS-assisted transmission requires derivation for the sum of the product of the near-field (depicted by $|g_i|$) and the far-field (represented by $|h_i^f|$) channel coefficient.
	
	Using $N=1$ in \eqref{eq:snr1}, we can get SNR for the single-element RRS system as 
	\begin{flalign}\label{eq:snr2}
		\gamma = \bar{\gamma} \left( |g_i| |h_i^f| \right)^2
	\end{flalign}

The specific case illustrated in \eqref{eq:snr2} simplifies the analysis of the RRS system.
	\subsection{Channel Models}
	There are many  small-scale fading models  models such as Rayleigh, Rice, Nakagami-m, $\alpha$-$\mu$, $\kappa$-$\mu$, and $\eta$-$\mu$ which may not provide enough flexibility to accurately  fit the measurement data in some  propagation scenarios, requiring more generalized and flexible models. The authors in \cite{Marins2019_alpha_eta_kapp_mu} demonstrated that the $\alpha$-$\eta$-$\kappa$-$\mu$  is the best option fitting over a wide range of situations, including intricate and rare events. This model includes almost all short-term propagation phenomena  to generalize the fading model for a wireless channel. The fading coefficient  $|h_i^f|$  distributed to $ \alpha $-$ \eta $-$ \kappa $-$ \mu $  is given by  \cite{Yacoub_2016_alpha_eta_kappa_mu}\cite{Silva_2020_alpha_eta_kappa_mu}:
	\begin{flalign} \label{eq:Ralpha}
	{|h_i^f|}^\alpha = \sum_{j=1}^{\mu_x} (X_j+{\lambda_{x_j}})^2 + \sum_{j=1}^{\mu_y} (Y_j+{\lambda_{y_j}})^2
	\end{flalign}
	where $\alpha$ denotes the non-linearity of the channel, $\mu_x$ and $\mu_y$ denote the number of multi-path clusters of in-phase component and quadrature component, respectively. $\lambda_{x_j}$ and $\lambda_{y_j}$ are the average values of the in-phase and quadrature components of the multi-path waves of the $j$-th cluster, respectively, and $X_j\sim {\cal{N}} (0, \sigma_x^2)$ and $Y_j\sim {\cal{N}} (0,\sigma_y^2)$ are mutually independent Gaussian processes, where $\sigma_x^2$ and $\sigma_y^2$ are variances of in-phase and quadrature components of the multi-path waves, respectively. In general, the $ \alpha $-$ \eta $-$ \kappa $-$ \mu $ model is quantified by seven different parameters, namely $\alpha$, $\eta$, $\kappa$, $\mu$, $p$, $q$, and $\hat{r}$. To define these parameter, denote the power of in-phase $(x)$ and quadrature-phase $(y)$ components of dominant $(d)$ waves and scattered $ (s) $ waves as $P_{ab}$, where $a \in \{d,s\}$ and $b \in \{x,y\}$. Thus, we define the parameters as $\eta = \frac{P_{sx}}{P_{sy}}$,  $\kappa = \frac{P_{dx} + P_{dy}}{P_{sx} + P_{sy}}$, $\mu = \frac{\mu_x+\mu_y}{2}$, $p = \frac{\mu_x}{\mu_y}$, $q = \frac{P_{dx}}{P_{dy}}/\frac{P_{sx}}{P_{sy}}$, and $\hat{r}= \sqrt[\alpha]{\mathbb{E}[R^{\alpha}]}$.
	
	For statistical analysis of the RRS system in \eqref{eq:snr1}, the density and distribution functions of the channel fading are required. Using \eqref{eq:Ralpha}, the authors in \cite{Silva_2020_alpha_eta_kappa_mu} presented the PDF $f_{|h_i^f|}(x)$ of the envelope $|h_i^f|$ in terms of the generalized Laguerre polynomial ($L_n$) and  the regularized hypergeometric function (${}_0\tilde{F}_1 $):
	\begin{flalign} \label{eq:pdf_series}
	f_{|h_i^f|}(x) &= \frac{\alpha (\xi \mu)^\mu}{\exp\Big(\frac{(1+pq)\kappa\mu}{\delta}\Big)} \bigg(\frac{p}{\eta}\bigg)^{\frac{p\mu}{1+p}} \frac{x^{\alpha\mu-1}}{\hat{r}^{\alpha \mu} } \exp\Big(-\frac{x^\alpha p \xi \mu}{\hat{r}^\alpha \eta}\Big) \nonumber \\ & \times \sum_{n=0}^{\infty} \Big(\frac{x^\alpha \xi \mu (p-\eta)}{\hat{r}^\alpha \eta}\Big)^n L_n^{\frac{\mu}{1+p}-1} \Big(\frac{\eta \kappa \mu}{\delta (\eta-p)}\Big) \nonumber \\ & \times {}_0\tilde{F}_1 \Big(;\mu+n;\frac{p^2qx^\alpha \kappa \xi \mu^2}{\hat{r}^\alpha \delta \eta}\Big) 
	\end{flalign}
	where $\xi = \frac{(1+\eta)(1+\kappa)}{(1+p)}$ and $\delta = \frac{(1+q\eta)(1+p)}{(1+\eta)}$.
	
	It can be seen that the PDF $f_{|h_i^f|}(x)$ contains an infinite series representation, which approximates the system performance when a finite number of terms are used for the convergence of the distribution function. It is always desirable to express the statistics of the channel fading in an exact form using tractable mathematical functions for efficient performance analysis and numerical computations. Further, analysis for RRS-based wireless system requires the PDF of the sum of random variables distributed according to  $ \alpha $-$ \eta $-$ \kappa $-$ \mu $.

\section{Statistical Results for RRS-Based Transmission}
In this section, we present exact formulation of the density and distribution functions for the resultant fading channel in RRS transmissions, incorporating the effects of both near-field and far-field propagation. This formulation leverages the multivariate Fox's H-function to accurately characterize the channel statistics, providing a unified and tractable mathematical framework for analyzing RRS-based systems under the $\alpha$-$\eta$-$\kappa$-$\mu$ fading model.

Using \eqref{eq:received equation1}, the resultant channel for an $N$-element RRS system is expressed as $
Z = \sum_{i=1}^N Z_i$, where $Z_i = |g_i| |h_i^f|$ represents the channel gain for a single-element RRS system. In this formulation, $|h_i^{f}|$ denotes the far-field channel coefficient, while $g_i$ corresponds to the near-field coefficient.

The approach leverages the moment generating function (MGF) technique to facilitate the statistical analysis of the resultant channel $Z$. Specifically, the MGF of $Z$ is expressed as the product of the MGFs of its independent components $Z_i$. By employing this formulation, the PDF of $Z$ is obtained through the inverse Laplace transform. The resulting expression for the PDF is given by 
\begin{flalign}\label{eqn:pdf_mgf_main}
f_{Z}(x) = \frac{1}{2\pi \J} \int_{\mathcal{L}} e^{tx} \prod_{i=1}^{N} M_{Z_i}(t) dt
\end{flalign}
where $\mathcal{L}$ represents the contour of integration in the complex plane, $ M_{Z_i}(t)$ denotes the MGF of $ Z_i$, and $ e^{tx} $ facilitates the transformation back to the PDF domain. This formulation provides a structured way for deriving the PDF of the resultant channel $Z$ in an $N$-element RRS system.

\subsection{Infinite-Series Representation for RRS Channel $Z$}
In this subsection, we use the existing representation of the $\alpha$-$\eta$-$\kappa$-$\mu$ fading model $|h_i^{f}|$  containing infinite series of \cite{Silva_2020_alpha_eta_kappa_mu} as presented in \eqref{eq:pdf_series} to develop the PDF of the resultant channel $Z$ for an $N$-element RRS-based transmission. 

	\begin{my_theorem} \label{th:mgf_infinity}
The PDF of the resultant channel in an $N$-element RRS-based transmission, based on the existing $\alpha$-$\eta$-$\kappa$-$\mu$ fading model representation, is expressed as an $N$-infinite series sum of $2N$-variate Fox-H functions:
		\begin{flalign}\label{eq:mgf_infinity}
			&	f_{Z}(x) =  \sum_{n_1=0}^{\infty} \sum_{n_2=0}^{\infty} \sum_{n_3=0}^{\infty} \cdots \sum_{n_N=0}^{\infty} x ^{ \sum_{i=1}^N(\alpha_i n_i -1)}  \nonumber \\&\times \prod_{j=1}^{N} \frac{\alpha_j (\xi_j \mu_j)^{\mu_j}}{(g_j \hat{r}_j)^{\alpha_j \mu_j} \exp\left( \frac{(1 + p_j q_j) \kappa_j \mu_j}{\delta_j} \right)}\nonumber \\& \times\left( \frac{p_j}{\eta_j} \right)^{\frac{p_j \mu_j}{1 + p_j}} \left( \frac{\xi_j \mu_j (p_j - \eta_j)}{(g_j \hat{r}_j)^{\alpha_j} \eta_j} \right)^n
			L_n^{\frac{\mu_j}{1 + p_j} - 1} \left( \frac{\eta_j \kappa_j \mu_j}{\delta_j (\eta_j - p_j)} \right)\nonumber \\&\times
			H^{{0,1}:\{1,0;1,0\}_N}_{{1,1}:\{0,1;0,2\}_N}  \Bigg[ \begin{matrix}~ V_1~ \\ ~V_2~ \end{matrix} \Bigg| \begin{matrix}~\{\zeta_{1,i}\}_{i=1}^	N \\ \{\zeta_{2,i}\}_{i=1}^	N  \end{matrix}\Bigg], 	
		\end{flalign}
		where $\zeta_{1,i}=\left( \frac{p_i \xi_i \mu_i}{( g_i \hat{r}_i)^{\alpha_i} \eta_i} x ^{\alpha_i}\right)$ and $\zeta_{2,i}=	\left( \frac{p_i^2 q_i \kappa_i \xi_i \mu_i^2}{( g_i \hat{r}_i)^{\alpha_i} \delta_i \eta_i} x ^{\alpha_i}\right)$; $V_1=\{(1-\sum_{i=1}^{N} (\mu_i+n);\{ \alpha_i\}_{i=1}^N,\{ \alpha_i\}_{i=1}^N):(-,-);  (-,-)\}$ and
		$V_2=\{(1-\sum_{i=1}^{N} \alpha_i;\{ \alpha_i\}_{i=1}^N) ,\{\alpha_i\}_{i=1}^N:(0,1);(0,1),(1-\mu_i-\eta_i)\}$	

	\end{my_theorem}

	\begin{IEEEproof}
	The proof is presented in Appendix A.
\end{IEEEproof}
The PDF in \eqref{eq:mgf_infinity} is computationally intensive, as it involves an 
$N$-term infinite series over $2N$-variate Fox-$H$ functions. Although it is not feasible to express the sum for i.ni.d.
$\alpha$-$\eta$-$\kappa$-$\mu$ variate using standard mathematical functions, multivariate Fox-$H$ function can provide exact and asymptotic performance of wireless systems under generalized fading models.  Recently,  Fox's H functions are finding applications in wireless communication research unifying performance analysis for intricate  fading distributions. However, infinite-sum representation should be avoided	since it  approximate the system performance when a finite number of terms are used for the convergence of the distribution function.

 \subsection{Exact Representation for RRS Channel $Z$ }	
 We introduce a novel approach that derives an exact statistical representation of the  $\alpha$-$\eta$-$\kappa$-$\mu$ fading model. This approach expresses  PDF of the channel envelope $|h_i^f|$ using a single trivariate Fox-$H$ function, avoiding the need for infinite series approximations. First, we derive exact expressions of the PDF for  the channel envelope $|h_i^f|$ for the single-element RRS  distributed according to the $\alpha$-$\eta$-$\kappa$-$\mu$ model using a single Fox's H-function. Next, we utilize the statistical results of  $Z_i$ to analyze the $N$-element RRS-based transmission  by expressing the sum of trivariate Fox-H functions in terms of a $3N$-variate Fox-H function for $Z$.  

We define $ \psi_1 = \frac{p\alpha\mu^{2}\xi^{1+\frac{\mu}{2}}\delta^{\frac{\mu}{2}-1}q^{\frac{1+p-p\mu}{2+2p}}\eta^{-\frac{1+p+p\mu}{2+2p}}}{\kappa^{\frac{\mu}{2}-1}\exp \left (\frac{(1+pq)\kappa\mu}{\delta}\right)}$, $ \psi_2$ = $\alpha-1$, $ \psi_3= \frac{p\xi\mu}{\eta\hat{r}^\alpha} $, $ A_{1} = \frac{p\mu}{1+p}$-$1, A_{2}= \frac{\mu}{1+p}$-$1, A_{3}= \frac{(\eta-p)\xi\mu}{\eta\hat{r}^\alpha} $, $ A_{4} = 2p\mu \sqrt{\frac{q\kappa\xi}{\eta\delta\hat{r}^\alpha}} $, and $A_{5} = 2\mu\sqrt{\frac{\kappa\xi}{\delta\hat{r}^\alpha}} $.

Defining $U = \sum_{i=1}^{\mu_x} (X_i+\lambda_{x_i})^2$ and $V = \sum_{i=1}^{\mu_y} (X_i+\lambda_{y_i})^2$ in \eqref{eq:Ralpha}, we represent $R^\alpha = U+V$. 
By means of random variable transformation, the PDF $f_|h_i^f|(x)$ of $ |h_i^f| $ can be obtained as
\begin{flalign} \label{eq:derivation_1}
	f_{|h_i^f|}(x) = \alpha x^{\alpha-1} \int_{0}^{x^\alpha} f_{U}(x^\alpha-v) f_{V}(v) dv
\end{flalign}
In the following lemma, we employ the Fox-H function to solve \eqref{eq:derivation_1}, deriving the PDF of the channel envelope for the $\alpha$-$\eta$-$\kappa$-$\mu$ fading model without infinite series.
	\begin{my_lemma}
		The PDF of the channel envelope for the  $\alpha$-$\eta$-$\kappa$-$\mu$ fading model is given by
		\begin{flalign} \label{eq:pdf_new_without_pe}
			&f_{|h_i^f|}(x) = \frac{\psi_1 \pi^2 2^{(2-\mu)} A_4^{A_1} A_5^{A_2} x^{\alpha\mu-1} e^{-\psi_3 x^\alpha}}{(\hat{r}^\alpha)^{1+\frac{\mu}{2}}} \nonumber \\ &\times H^{0,1;1,0;1,1;1,0}_{1,1;0,1;2,3;1,3} \Bigg[ \begin{matrix}~ V_1~ \\ ~V_2~ \end{matrix} \Bigg| A_{3}x^{\alpha}, \frac{A_4^2}{4}x^{\alpha},\frac{A_5^2}{4}x^{\alpha}\Bigg],
		\end{flalign}	
		where $V_1 = \big\{(-A_2;1,0,1)\big\}: \big\{(-,-)\big\} ; \big\{(-A_1,1)(\frac{1}{2},1)\big\} ;\big\{(\frac{1}{2},1)\big\} $ and $V_2 = \big\{(-1-A_1 -A_2;1,1,1) \big\} : \big\{(0,1) \big\} ; \big\{(0,1),(-A_1,1),(\frac{1}{2},1) \big\} ; \big\{(0,1),(-A_2,1),(\frac{1}{2},1) \big\}$.
	\end{my_lemma}
	
	\begin{IEEEproof}	
		See Appendix B.
	\end{IEEEproof}
	
It is evident that the PDF in \eqref{eq:pdf_new_without_pe} involves a single tri-variate Fox-H function without any infinite series, unlike \eqref{eq:pdf_series}. The use of the Fox-H function facilitates a more straightforward asymptotic analysis of the PDF in \eqref{eq:pdf_new_without_pe} compared to the infinite-series representation in \eqref{eq:pdf_series}.
The asymptotic expansion of the Fox-H function may exhibit superior properties, offering more accurate approximations across a broader range of parameters:
	\begin{my_proposition}
		An asymptotic expression for PDF of the $\alpha$-$\eta$-$\kappa$-$\mu$ fading channel in \eqref{eq:pdf_new_without_pe} is given by 
		\begin{flalign} \label{eq:asympt_pdf}
			\lim_{x\to0}f_{|h_i^f|}(x) = & \frac{\psi_1 \pi^2 2^{(2-\mu)} A_4^{A_1} A_5^{A_2} x^{\alpha\mu-1} }{4(\hat{r}^\alpha)^{1+\frac{\mu}{2}} e^{\psi_3 x^\alpha}} \frac{1}{\Gamma(\mu) \Gamma(1+A_2) \Gamma(\frac{1}{2})}
		\end{flalign}	
	\end{my_proposition}
	
	\begin{IEEEproof}
		The existing literature derives the asymptotic expression by computing the residue of dominant pole of a multi-variate Fox's H-function, which sometimes become tedious. We convert the tri-variate Fox's H-function using the  product of three Meijer's G-function to derive the asymptotic expression of the PDF  using simpler mathematical functions. Thus,  eliminating the linear combination terms of the tri-variate Fox's H-function, we can express the PDF in \eqref{eq:pdf_new_without_pe} as
	
		\begin{flalign}\label{eq:three_meijer}
			f_{|h_i^f|}(x) \approx & \frac{\psi_1 \pi^2 2^{(2-\mu)} A_4^{A_1} A_5^{A_2} x^{\alpha\mu-1} e^{-\psi_3 x^\alpha}}{(\hat{r}^\alpha)^{1+\frac{\mu}{2}}} G^{1,0}_{0,1}\bigg(\begin{matrix} - \\ 0 \end{matrix} \bigg| A_3 x^\alpha \bigg) \nonumber \\ & \times  G^{1,1}_{2,4}\bigg(\begin{matrix} -A1, \frac{1}{2} \\ 0, -A_1, \frac{1}{2}, -1-A_1-A_2 \end{matrix} \bigg| \frac{A_4^2 x^\alpha}{4} \bigg)  \nonumber \\ & \times   G^{1,0}_{1,3}\bigg(\begin{matrix} \frac{1}{2} \\ 0, -A_2, \frac{1}{2} \end{matrix} \bigg| \frac{A_5^2 x^\alpha}{4} \bigg)
		\end{flalign} 
			
		Applying \cite{Meijers_asymp_low_snr}  for the asymptotic expansion at $x \to 0$ for Meijer's G-functions in \eqref{eq:three_meijer}, leads to the asymptotic PDF in \eqref{eq:asympt_pdf}.
	\end{IEEEproof}
	It should be mentioned that the authors in \cite{Silva_2020_alpha_eta_kappa_mu} provided  asymptotic expression for the PDF of the  $\alpha$-$\eta$-$\kappa$-$\mu$ distribution without any series expression.  However, as demonstrated through simulations in the next Section,  the proposed asymptotic PDF in  \eqref{eq:asympt_pdf} offers a more accurate  representation than \cite{Silva_2020_alpha_eta_kappa_mu}.

In the following theorem, we build on the statistical results of the single-element RRS in \eqref{eq:pdf_new_without_pe} to derive the PDF of the $N$-element RRS-based transmission, denoted as $Z$, expressed in terms of the multivariate Fox-H function.

		\begin{my_theorem} \label{th:pdf_mrc}
		The PDF of the resultant channel for an $N$-element RRS-based transmission is expressed as  $3N$-variate/multivariate Fox-H function :
		
		\begin{flalign}\label{eq:mrc_pdf_foxh}
		&	f_{Z}(x) =   x ^{ \sum_{i=1}^N(\alpha_i \mu_i)} \prod_{j=1}^{N}|g_j|^{-\alpha\mu_j+1} \frac{\psi_1}{(\hat{r}^{\alpha_j})^{1+\frac{\mu_j}{2}}} \pi^2 2^{(2-\mu_j)}\nonumber \\&\times A_{j,4}^{A_{j,1}} A_{j,5}^{A_{j,2}}H^{0,2:\{1,0;1,0;1,0;1,0\}_N}_{2,2:\{0,1;1,2;1,2;0,1\}_N} \nonumber \\& \times\Bigg[ \begin{matrix}~ V_1~ \\ ~V_2~ \end{matrix} \bigg|\{\zeta_{i,1}\}_{i=1}^	N , \{\zeta_{i,2}\}_{i=1}^N , \{\zeta_{i,3}\}_{i=1}^	N   \{\zeta_{i,4}\}_{i=1}^	N\Bigg], 	
	\end{flalign}
	where
		$V_1 = \big\{(1-\sum_{i=1}^{N}\alpha_i\mu_i;\{\alpha_i,\alpha_i,\alpha_i,\alpha_i\})(-{A_{2,i}};1,0,1,0): -;(\frac{1}{2},1);(\frac{1}{2},1) ;- $  and $V_2 = \big\{(1-\sum_{i=1}^{N}\alpha_i\mu_i;(\alpha_i,\alpha_i,\alpha_i,\alpha_i)_{i=1}^N)(-1-{A_{1,i}} -{A_{2,i}};1,1,1,0):(0,1) ;(0,1)(\frac{1}{2},1);(0,1)(-A_{i,2},1)(\frac{1}{2},1);(0,1)\big\}$.
		 $\zeta_{i,1}=\left( A_{i,3} |g_i|^{-\alpha_i}x^{\alpha_i}\right)^{s_{i,1}}$ , $\zeta_{i,2}=	\left(\frac{( A_{i,4}^2 |g_i|^{-\alpha_i}x^{\alpha_i}}{4}\right)^{s_{i,2}}$ , $\zeta_{i,3}=	\left(\frac{( A_{i,5}^2 |g_i|^{-\alpha_i}x^{\alpha_i}}{4}\right)^{s_{i,3}}$ and   $\zeta_{i,4}=	\left(\psi_3 |g_i|^{-\alpha_i}x^{\alpha_i}\right)^{s_{i,4}}$
	
	\end{my_theorem}
	\begin{IEEEproof}
	The proof is presented in Appendix C.
	\end{IEEEproof}
Comparing \eqref{eq:mgf_infinity} and \eqref{eq:mrc_pdf_foxh}, it can be seen that the  use of multivariate Fox-H function representation for  PDF is inevitable due to mathematical intricacy of the $\alpha$-$\eta$-$\kappa$-$\mu$ distribution. The Fox-H function is increasingly popular in wireless communication research, as it provides a unified framework for performance analysis across complex fading distributions.  However, \eqref{eq:mrc_pdf_foxh} does not contain any infinite-series which may provide better statistical performance assessment over generalized fading channels.
  
In what follows, we use the exact statistical analysis in \eqref{eq:pdf_new_without_pe} and \eqref{eq:mrc_pdf_foxh}  to analyze the outage probability and average BER performance for single-element $N=1$ and multiple-element $N\geq 1$ RRS systems.
	\section{Performance Analysis for RRS-Based Systems }
An accurate performance evaluation of RRS-based transmission over fading channels is crucial for optimizing deployment scenarios in practical setups. Leveraging the results from Lemma 1 and Theorem 2, we analyze the performance of single-element and multiple-element RRS systems operating in an $\alpha$-$\eta$-$\kappa$-$\mu$ fading channel. Metrics such as outage probability and average BER are employed to provide a detailed understanding of the system's behaviour under varying channel conditions. It is worth mentioning that the ergodic capacity of RRS systems can also be derived using a similar approach; however, it is omitted here to avoid redundancy.

Let $\gamma$ represent the instantaneous SNR and $\gamma_{\rm th}$ denote the threshold SNR, which serves as a quality-of-service metric. The outage probability of a communication system is defined as 
\begin{flalign}\label{eq:OP}
P_{\rm out} = \Pr(\gamma < \gamma_{\rm th}) = F_{\gamma}(\gamma_{\rm th}),
\end{flalign}
where $ F_{\gamma}(\gamma_{\rm th})$ represents the cumulative distribution function (CDF) of the SNR $\gamma$ evaluated at $\gamma_{\rm th}$.

	Average BER for binary modulations can be expressed using the CDF \cite{Ansari2011}:
\begin{flalign} \label{eq:ber_eqn}
	\overline{P_e} = \frac{ q_{\rm m}^{p_{\rm m}}}{2\Gamma(p_m)} \int_{0}^{\infty} \gamma^{p_m-1} e^{-q_m\gamma} F_\gamma(\gamma) d\gamma,
\end{flalign}
where $p_m$ and $q_m$ determine the type of modulation scheme used. Thus, binary phase shift keying (BPSK) can be represented by $\{ p_m  = 0.5, q_m  = 1\}$, while differential phase shift keying (DPSK) and binary frequency shift keying (BFSK) are characterized by $ \{p_m = 1, q_m = 1 \}$, and $ \{p_m = 0.5, q_m = 0.5\} $, respectively.

	In the following subsections, we use the derived statistical results to present the exact analysis of the outage probability and average BER of a wireless link subjected to $\alpha$-$\eta$-$\kappa$-$\mu$ fading.

	\subsection{Single-Element RRS System}
To analyze the performance of single-element RRS system, we require the PDF and CDF of the SNR $\gamma$ as depicted in \eqref{eq:snr2}. Using standard transformation of random variable, the PDF of SNR is given as 
	$f_\gamma(\gamma) = \frac{1}{2g_i\sqrt{\gamma \bar{\gamma}}}f_{|h_i^f|}(\frac{1}{g_i}\sqrt{\frac{\gamma}{\bar{\gamma}}})$, where $f_{|h_i^f|}(\cdot)$ is given in \eqref{eq:pdf_new_without_pe}.
	
	\subsubsection{Outage Probability}
 Using \eqref{eq:pdf_new_without_pe} in $P_{\rm out} = \Pr(\gamma < \gamma_{\rm th}) = \int_{0}^{\gamma_{\rm th}}\frac{1}{2g_i\sqrt{\gamma \bar{\gamma}}}f_{|h_i^f|}(\frac{1}{g_i}\sqrt{\frac{\gamma}{\bar{\gamma}}}) d\gamma$ and applying the standard procedure for multiple Mellin-Barnes integrals, we represent  the outage probability as
	\begin{flalign} \label{eq:outage_new_without_pe}
		&P_{\rm out} =F_\gamma(\gamma_{\rm th})= \frac{\psi_1  \pi^2 2^{(2-\mu)} A_4^{A_1} A_5^{A_2} \gamma_{\rm th}^{\frac{\alpha\mu}{2}}}{(\hat{r}^\alpha)^{1+\frac{\mu}{2}} \bar{\gamma}^{\frac{\alpha\mu}{2}} g_i^{\alpha \mu}} \nonumber \\ &\times H^{0,2;1,0;1,1;1,0;1,0}_{2,2;0,1;2,3;1,3;0,1} \Bigg[\begin{matrix}~ V_3~ \\ ~V_4~ \end{matrix} \Bigg| \frac{A_{3}\gamma_{\rm th}^{\frac{\alpha}{2}}}{\bar{\gamma}^{\frac{\alpha}{2}} g_i^{\alpha \mu}}, \frac{{A_{4}^{2}}\gamma_{\rm th}^{\frac{\alpha}{2}}}{4\bar{\gamma}^{\frac{\alpha}{2}} g_i^{\alpha \mu}},\frac{{A_{5}^{2}}\gamma_{\rm th}^{\frac{\alpha}{2}}}{4 \bar{\gamma}^{\frac{\alpha}{2}} g_i^{\alpha \mu}}, \frac{\psi_3 \gamma_{\rm th}^{\frac{\alpha}{2}}}{\bar{\gamma}^{\frac{\alpha}{2}} g_i^{\alpha \mu}} \Bigg],
	\end{flalign}	
where $V_3 = \big\{(-A_2;1,0,1,0),(1-{\alpha\mu}; \alpha,\alpha,\alpha,\alpha)\big\}: \big\{(-,-)\big\} ; \big\{(-A_1,1)(\frac{1}{2},1)\big\} ; \big\{(\frac{1}{2},1)\big\} ; \big\{(-,-)\big\} $ and $V_4 = \big\{(-1-A_1-A_2;1,1,1,0),(-{\alpha\mu};\alpha,\alpha,\alpha,\alpha) \big\} : \big\{(0,1) \big\} ; \big\{(0,1),(-A_1,1),(\frac{1}{2},1) \big\} ; \big\{(0,1),(-A_2,1),(\frac{1}{2},1) \big\} ; \\ \big\{(0,1)\big\}$.	

The  outage probability in a high SNR regime can provide useful insight for system design. We compute the residue at dominant poles to derive an  asymptotic expression for the outage probability:
	\begin{flalign}
		P_{\rm out}^\infty =& \frac{\psi_1 2^{(2-\mu)} A_4^{A_1} A_5^{A_2} \gamma_{\rm th}^{\frac{\alpha\mu}{2}}} { (\hat{r}^\alpha)^{1+\frac{\mu}{2}} \bar{\gamma}^{\frac{\alpha\mu}{2}}  g_i^{\alpha \mu}} \frac{1}{\Gamma(\mu) \Gamma(A_2)} .
	\end{flalign}
	The asymptotic outage can be represented as $P_{\rm out}^\infty = G_c \bar{\gamma}^{G_d}$, where $G_c$ and $G_d$ denote  coding gain and diversity order, respectively. Observing the exponent of average SNR $\bar{\gamma}$, we can deduce that the diversity order $G_d$ of the system over the $\alpha$-$\eta$-$\kappa$-$\mu$ fading channel is $\frac{\alpha\mu}{2}$. Thus, the other parameters, such as $\eta$ and $\kappa$, affect the system's coding gain $G_c$.

	\subsubsection{Average BER}
Substituting \eqref{eq:outage_new_without_pe} with $\gamma_{\rm th}=\gamma$ in \eqref{eq:ber_eqn}, using the integral form of Fox's H-function, and changing the order of integration, we get
	
	\begin{flalign} \label{eq:ber_derive}
		& \overline{P_e} =  \frac{\psi_1  \pi^2 2^{(2-\mu)} A_4^{A_1} A_5^{A_2}  q_m^{p_m}}{2\Gamma(p_m)\bar{\gamma}^{\frac{\alpha\mu}{2}} g_i^{\alpha \mu}} \Big(\frac{1}{2\pi \J}\Big)^{4} \nonumber 
		\\ & \times \int_{\mathcal{L}_1}\int_{\mathcal{L}_2}\int_{\mathcal{L}_3} \int_{\mathcal{L}_4} \frac{\Gamma(-s_1) \Gamma(-s_2)\Gamma(1+{A_1}+s_2)}{\Gamma(1+A_1+s_2)\Gamma(\frac{1}{2}+s_2)\Gamma(\frac{1}{2}-s_2)}
		\nonumber \\ &  \times \frac{\Gamma(-s_3)}{\Gamma(1+A_2+s_3)\Gamma(1/2+s_3)\Gamma(1/2-s_3)} \nonumber \\ & \times
		\frac{\Gamma(1+{A_2}+s_1+s_3)}{\Gamma(2+{A_2}+s_1+s_3+{A_1}+s_2)} \nonumber \\ & \times \frac{\Gamma(\alpha(s_1+s_2+s_3+s_4)+{\alpha\mu})}{\Gamma(\alpha(s_1+s_2+s_3+s_4)+{\alpha\mu}+1)} \nonumber \\ & \times
		\Big(\frac{A_{3}}{\bar{\gamma}^\frac{\alpha}{2} g_i^{\alpha \mu}}\Big)^{s_1} \Big(\frac{{A_{4}^{2}}}{4\bar{\gamma}^{\frac{\alpha}{2}} g_i^{\alpha \mu}}\Big)^{s_2}   \Big(\frac{{A_{5}^{2}}}{4\bar{\gamma}^{\frac{\alpha}{2}} g_i^{\alpha \mu}}\Big)^{s_3}\nonumber \\& \Big(\frac{\psi_3}{\bar{\gamma}^\frac{\alpha}{2} g_i^{\alpha \mu}}\Big)^{s_4} I_3 ds_1 ds_2 ds_3 ds_4 
	\end{flalign}
	where $\mathcal{L}_1$, $\mathcal{L}_2$, and $\mathcal{L}_3$, and $\mathcal{L}_4$ denote the contour in the complex plane. The inner integral $I_3$ can be simplified using \cite[(3.381/4)]{Gradshteyn} to 
	
	\begin{flalign} 
		I_3 &= \int_{0}^{\infty} \gamma^{p_m-1} e^{-q_m\gamma} \gamma^{\frac{\alpha}{2}(s_1+s_2+s_3+s_4)+\frac{\alpha\mu}{2}}  d\gamma \nonumber 
		\\ &   = q_m^{-(\frac{\alpha}{2}(s_1+s_2+s_3+s_4)+\frac{\alpha\mu}{2})-p_m} \nonumber \\ &  \times \Gamma(\frac{\alpha}{2}(s_1+s_2+s_3+s_4)+\frac{\alpha\mu}{2}+p_m)
	\end{flalign} 
	Substituting $I_3$ in \eqref{eq:ber_derive} and applying the definition of multivariate Fox's H-function, we get the average BER for the considered system as
	\begin{flalign} \label{eq:ber_new_without_pe}
		&\overline{P_e} = \frac{\psi_1 \pi^2 2^{(2-\mu)} A_4^{A_1} A_5^{A_2} }{2\Gamma(p_m) \bar{\gamma}^{\frac{\alpha\mu}{2}} q_m^{\frac{\alpha\mu}{2}}g_i^{\alpha \mu}} H^{0,3;1,0;1,1;1,0;1,0}_{3,2;0,1;2,3;1,3;0,1} \nonumber \\ & \Bigg[\begin{matrix} ~V_3~ \\ ~V_4~ \end{matrix} \Bigg| \frac{A_{3}}{\bar{\gamma}^{\frac{\alpha}{2}} g_i^{\alpha \mu}q_m^{\frac{\alpha}{2}}}, \frac{{A_{4}^{2}}}{4\bar{\gamma}^{\frac{\alpha}{2}}  g_i^{\alpha \mu}q_m^{\frac{\alpha}{2}}},\frac{{A_{5}^{2}}}{4\bar{\gamma}^{\frac{\alpha}{2}}  g_i^{\alpha \mu} q_m^{\frac{\alpha}{2}}}, \frac{\psi_3}{\bar{\gamma}^{\frac{\alpha}{2}}  g_i^{\alpha \mu} q_m^{\frac{\alpha}{2}}} \Bigg],
	\end{flalign}	
	where $V_3 = \big\{(-{A_2};1,0,1,0),(1-{\alpha\mu}; \alpha,\alpha,\alpha,\alpha), (1-\frac{\alpha\mu}{2}-p_m; \frac{\alpha}{2},\frac{\alpha}{2},\frac{\alpha}{2},\frac{\alpha}{2})\big\}: \big\{(-,-)\big\} ; \big\{(-{A_1},1)(\frac{1}{2},1)\big\} ; \big\{(\frac{1}{2},1)\big\} ; \big\{(-,-)\big\} $ and $V_4 = \big\{(-1-{A_1} -{A_2};1,1,1,0),(-\alpha^2\mu;\alpha,\alpha,\alpha,\alpha) \big\} : \big\{(0,1) \big\} ; \big\{(0,1),(-A_1,1),(\frac{1}{2},1) \big\} ; \big\{(0,1),(-A_2,1),(\frac{1}{2},1) \big\} ; \\ \big\{(0,1)\big\}$.

 \subsection{Multiple-Element RRS System}
The statistical performance analysis of a multi-element RRS system depends on  the SNR, as defined in \eqref{eq:snr1}. This involves deriving the sum of the products of near-field and far-field channel coefficients, represented as \( Z = \sum_{i=1}^N Z_i \), where \( Z_i = |g_i| |h_i^f| \). In this subsection, we employ the exact PDF of \( Z \), as presented in \eqref{eq:mrc_pdf_foxh} (see Theorem 2), to derive the outage probability and the average BER performance of the multiple-element RRS system.
 \subsubsection{Outage Probability}
 Using \eqref{eq:mrc_pdf_foxh} in $P_{\rm out} = Pr(\gamma < \gamma_{\rm th}) =\int_{0}^{\gamma_{\rm th}}\frac{1}{2\sqrt{\gamma \bar{\gamma}}}f_Z(\sqrt{\frac{\gamma}{\bar{\gamma}}}) d\gamma$, we get
			\begin{flalign}
		&P_{\rm out}=F_{\gamma}(\gamma_{\rm th})= \frac{1}{2\sqrt{ \bar{\gamma}}}\prod_{j=1}^{N}|g_j|^{-\alpha\mu_j+1} \frac{\psi_1}{(\hat{r}^{\alpha_j})^{1+\frac{\mu_j}{2}}} \pi^2 2^{(2-\mu_j)} \nonumber \\ &\times A_{j,4}^{A_{j,1}} A_{j,5}^{A_{j,2}}    \frac{1}{(2\pi \J)^4}\int_{\mathcal{L}_{i,1}}\int_{\mathcal{L}_{i,2}}\int_{\mathcal{L}_{i,3}} \int_{\mathcal{L}_{i,4}} \cdots \int_{\mathcal{L}_{N,1}}\int_{\mathcal{L}_{N,2}}\int_{\mathcal{L}_{N,3}}\nonumber \\& \times  \int_{\mathcal{L}_{N,4}} \bar{\gamma}^\frac{-\sum_{i=1}^{N}{\alpha_i ( \mu_i +  s_{i,1} + s_{i,2} + s_{i,3} + s_{i,4})-1}}{2}  \nonumber \\& \times\frac{1}{\Gamma(\sum_{i=1}^{N}\alpha_i ( \mu_i +  s_{i,1} + s_{i,2} + s_{i,3} + s_{i,4}))}\\&  \times  \int_{0}^{\gamma_{\rm th}}\gamma^\frac{\sum_{i=1}^{N}{\alpha_i ( \mu_i +  s_{i,1} + s_{i,2} + s_{i,3} + s_{i,4})-1}}{2} d\gamma \nonumber \\&  \times \prod_{i=1}^{N}\Gamma(\alpha_i ( \mu_i +  s_{i,1} + s_{i,2} + s_{i,3} + s_{i,4})) \Gamma(-s_{i,1})\nonumber \\& \times({A_{i,3}|g_i|^{-\alpha_i}})^{s_{i,1}}   \frac{\Gamma(-s_{i,2})}{\Gamma(1+A_{i,1}+s_{i,2})\Gamma(\frac{1}{2}+s_{i,2})\Gamma(\frac{1}{2}-s_{i,2})}\nonumber
		\\& \times \Big(\frac{A_{i,4}^{2}|g_i|^{-\alpha_i}}{4} \Big)^{s_{i,2}}   \frac{\Gamma(-s_{i,3})}{\Gamma(1+A_{i,2}+s_{i,3})\Gamma(\frac{1}{2}+s_{i,3})\Gamma(\frac{1}{2}-s_{i,3})}\nonumber
		\\& \times\Big(\frac{A_{i,5}^{2}|g_i|^{-\alpha_i}}{4} \Big)^{s_{i,3}}\nonumber   \frac{\Gamma(1+A_{i,2}+s_{i,1}+s_{i,3})\Gamma(1+A_{i,1}+s_{i,2})}{\Gamma(2+A_{i,1}+A_{i,2}+s_{i,1}+s_{i,2}+s_{i,3})}\nonumber \\& \times  \Gamma(-s_{i,4}) \big(\psi_3|g_i|^{-\alpha_i}\big)^{s_{i,4}}  
		ds_{i,1} ds_{i,2}ds_{i,3}  ds_{i,4}  
	\end{flalign}
Solving the inner integral and applying the multivariate Fox-H definition, we get	

		\begin{flalign}\label{eq:pdf_mv}
		&P_{\rm out}= \prod_{j=1}^{N}|g_j|^{-\alpha\mu_j+1} \frac{\psi_1}{2(\hat{r}^{\alpha_j})^{1+\frac{\mu_j}{2}}} \pi^2 2^{(2-\mu_j)} A_{j,4}^{A_{j,1}}A_{j,5}^{A_{j,2}}\nonumber \\&\frac{\gamma_{th}^{\frac{\alpha_j\mu_j+1}{2}}}{\bar{\gamma}\frac{\alpha_j\mu_j+1}{2}}\frac{2}{\sum_{j=1}^{N}\alpha_j\mu_j+1}
		H^{0,3;\{1,0;1,0;1,0;1,0\}_{N}}_{3,3:\{0,1;1,2;1,2;0,1\}_{N}} \nonumber \\ & \Bigg[ \begin{matrix}~ V_1~ \\ ~V_2~ \end{matrix} \bigg|\{\zeta_{i,1}\}_{i=1}^	N , \{\zeta_{i,2}\}_{i=1}^	N , \{\zeta_{i,3}\}_{i=1}^	N   \{\zeta_{i,4}\}_{i=1}^	N \Bigg], 	
		\end{flalign}
		where
		$V_1 = \big\{(1-\sum_{i=1}^{N}\alpha_i\mu_i;\{\alpha_i,\alpha_i,\alpha_i,\alpha_i\})(-{A_{2,i}};1,0,1,0)\nonumber \\  (1;\frac{1}{2}(\alpha_i,\alpha_i,\alpha_i,\alpha_i)_{i=1}^N): (-);(\frac{1}{2},1)(-);(-) (\frac{1}{2},1)(-)\big\}$  and $V_2 = \big\{(1-\sum_{i=1}^{N}\alpha_i\mu_i;(\alpha_i,\alpha_i,\alpha_i,\alpha_i)_{i=1}^N);(-1-A_{i,1} -A_{i,2};1,1,1,0)(-;\frac{1}{2}(\alpha_i,\alpha_i,\alpha_i,\alpha_i)_{i=1}^N):(0,1) ;(0,1)(\frac{1}{2},1);(0,1)(-A_{i,2},1)(\frac{1}{2},1))\big\}$.

		$\zeta_{i,1}=\left( A_{i,3} |g_i|^{-\alpha_i}\left(\frac{\gamma_{th}}{\bar{\gamma}}\right)^{\alpha_i}\right)^{s_{i,1}}$ , $\zeta_{i,2}=	\left(\frac{( A_{i,4}^2 |g_i|^{-\alpha_i}}{4 }\left(\frac{\gamma_{th}}{\bar{\gamma}}\right)^{\alpha_i}\right)^{s_{i,2}}$ , $\zeta_{i,3}=	\left(\frac{( A_{i,5}^2 |g_i|^{-\alpha_i}}{4}\left(\frac{\gamma_{th}}{\bar{\gamma}}\right)^{\alpha_i}\right)^{s_{i,3}}$ and   $\zeta_{i,4}=	\left(\psi_3 |g_i|^{-\alpha_i}\left(\frac{\gamma_{th}}{\bar{\gamma}}\right)^{\alpha_i}\right)^{s_{i,4}}$
		
		The asymptotic outage probability in high SNR region can be obtained by calculating the residue of \eqref{eq:pdf_mv} at dominant poles \cite{AboRahama_2018} as
		
		\begin{flalign} \label{eq:asym_outage_multi_rrs}
			P_{\rm out}^\infty =  \prod_{i=1}^{N}C_i \bar{\gamma}^{\frac{-\alpha_i \mu_i}{2}}
		\end{flalign}
	where $C_i$ is coding gain of the individual RRS element. Observing the exponent of the average SNR in \eqref{eq:asym_outage_multi_rrs}, the diversity order for outage probability can be obtained as $\sum_{i=1}^{N} \frac{\alpha_i \mu_i}{2}$, which is consistent with the results available in the literature.

	\subsubsection{Average BER}
	Using $\gamma_{\rm th}=\gamma$ in \eqref{eq:pdf_mv}	and substituting in \eqref{eq:ber_eqn} with resulting inner integral as
		\begin{flalign*}
	&I_1=\int_{0}^{\infty}\gamma^{p_m-1}e^{-q_m\gamma}\frac{\gamma^{\frac{\sum_{i=1}^{N}\alpha_i(\mu_i+s_{i,1}+s_{i,2}+s_{i,3}+s_{i,4})+1}{2}}}{\bar{\gamma}^{\frac{\sum_{i=1}^{N}\alpha_i(\mu_i+s_{i,1}+s_{i,2}+s_{i,3}+s_{i,4})+1}{2}}} d\gamma \nonumber \\&  =	\frac{q_m^{-\left(p_m+\frac{\sum_{i=1}^{N}\alpha_i(\mu_i+s_{i,1}+s_{i,2}+s_{i,3}+s_{i,4})+1}{2}\right)}}{\bar{\gamma}^{\frac{\sum_{i=1}^{N}\alpha_i(\mu_i+s_{i,1}+s_{i,2}+s_{i,3}+s_{i,4})+1}{2}}}\nonumber\\& \times \Gamma(p_m+\frac{(\sum_{i=1}^{N}\alpha_i(\mu_i+s_{i,1}+s_{i,2}+s_{i,3}+s_{i,4})+1)}{2}
	\end{flalign*}
and applying the definition of multi-variate Fox-H function, we get	
		 	\begin{flalign} \label{eq:ber_eqn2}
	 		&\overline{P_e} = \frac{ q_{\rm m}^{p_{\rm m}}}{2\Gamma(p_m)} \prod_{j=1}^{N}|g_j|^{-\alpha\mu_j+1} \frac{\psi_1}{2(\hat{r}^{\alpha_j})^{1+\frac{\mu_j}{2}}} \pi^2 2^{(2-\mu_j)} A_{j,4}^{A_{j,1}} A_{j,5}^{A_{j,2}}\nonumber \\ & \frac{q_m^{(-p_m+\frac{\sum_{j=1}^{N}\alpha_j\mu_j+1}{2})}}{\bar{\gamma}^{\sum_{j=1}^{N}\frac{1}{2}(\alpha_j\mu_j+1)}}\frac{2}{\sum_{j=1}^{N}\alpha_j\mu_j+1} 
	 		H^{0,5;\{1,0;1,0;1,0;1,0\}_{N}}_{5,3:\{0,1;1,2;1,2;0,1\}_{N}}\nonumber \\ &  \Bigg[ \begin{matrix}~ V_1~ \\ ~V_2~ \end{matrix} \bigg|\{\zeta_{i,1}\}_{i=1}^	N , \{\zeta_{i,2}\}_{i=1}^	N , \{\zeta_{i,3}\}_{i=1}^	N   \{\zeta_{i,4}\}_{i=1}^	N\bigg| \Bigg], 	
	 		\end{flalign}
	 		where
	 		$V_1 = \big\{(1-p_m+\frac{\sum_{i=1}^{N}\alpha_i\mu_i+1}{2};\frac{1}{2}\{\alpha_i,\alpha_i,\alpha_i,\alpha_i\}_{i=1}^N)(-{A_{2,i}};1,0,1,0)\nonumber\\(1;\frac{1}{2}(\alpha_i,\alpha_i,\alpha_i,\alpha_i)_{i=1}^N)(1-\sum_{i=1}^{N}\alpha_i\mu_i;\alpha_i,\alpha_i,\alpha_i,\alpha_i): (-);(\frac{1}{2},1)(-);(-) (\frac{1}{2},1)(-)\big\}$  and $V_2 = \big\{(1-\sum_{i=1}^{N}\alpha_i\mu_i;(\alpha_i,\alpha_i,\alpha_i,\alpha_i)_{i=1}^N);(-1-A_{i,1} -A_{i,2};1,1,1,0)(-;\frac{1}{2}(\alpha_i,\alpha_i,\alpha_i,\alpha_i)_{i=1}^N):(0,1) ;(0,1)(\frac{1}{2},1);(0,1)(-A_{i,2},1)(\frac{1}{2},1)\big\}$.

	 		$\zeta_{i,1}=\left( A_{i,3} |g_i|^{-\alpha_i}\left(\frac{q_m}{\bar{\gamma}}\right)^{-\alpha_i} \right)^{s_{i,1}}$ , $\zeta_{i,2}=	\left(\frac{( A_{i,4}^2 |g_i|^{-\alpha_i}}{4 }\left(\frac{q_m}{\bar{\gamma}}\right)^{-\alpha_i}\right)^{s_{i,2}}$ , $\zeta_{i,3}=	\left(\frac{( A_{i,5}^2 |g_i|^{-\alpha_i}}{4}\left(\frac{q_m}{\bar{\gamma}}\right)^{-\alpha_i}\right)^{s_{i,3}}$and $\zeta_{i,4}=\left(\psi_3|g_i|^{-\alpha_i}\left(\frac{q_m}{\bar{\gamma}}\right)^{-\alpha_i}\right)^{s_{i,4}}$.
		\begin{figure*}[t]
		\centering
		\subfigure[PDF]{\includegraphics[scale=0.28]{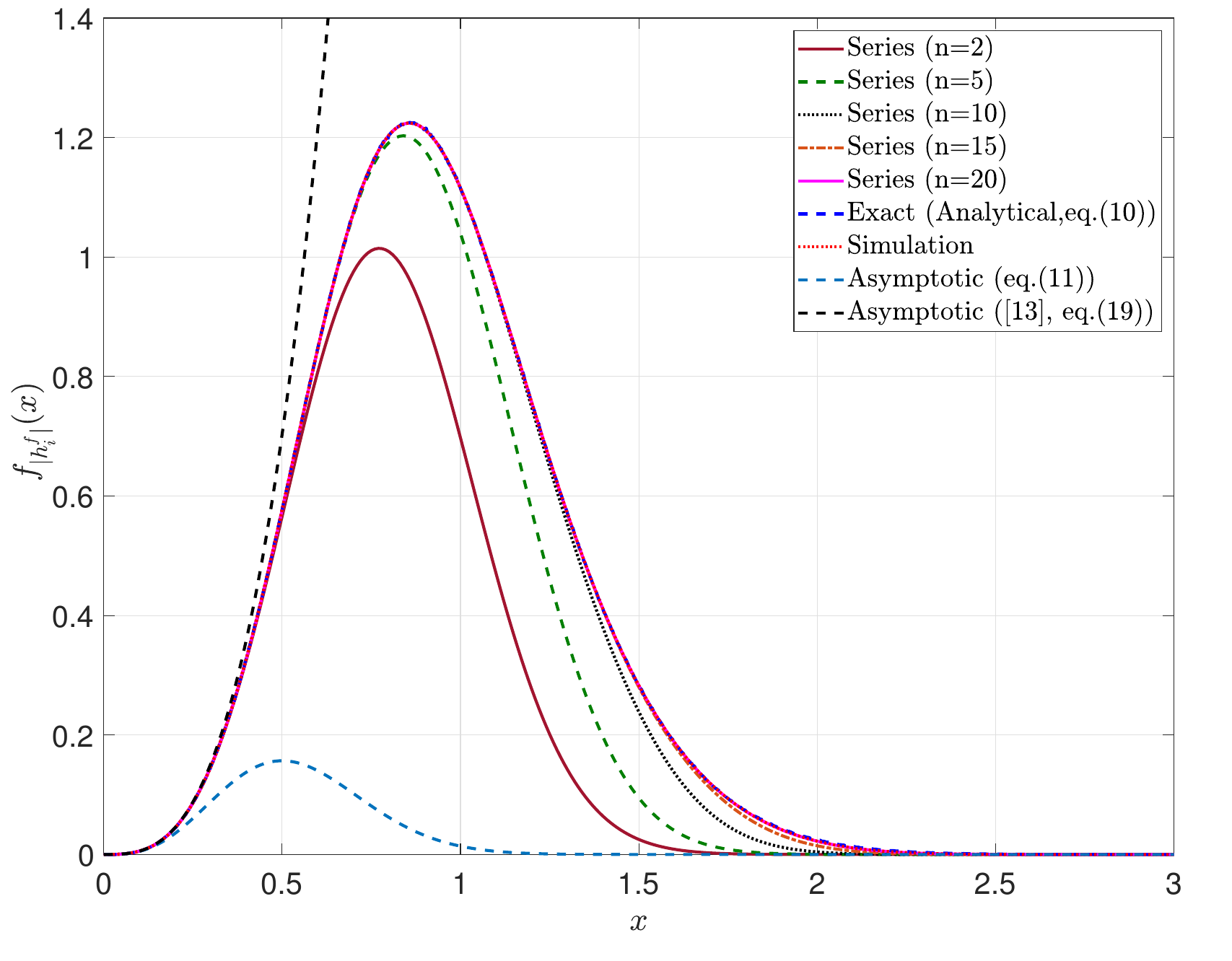}} 
		\subfigure[CDF]{\includegraphics[scale=0.28]{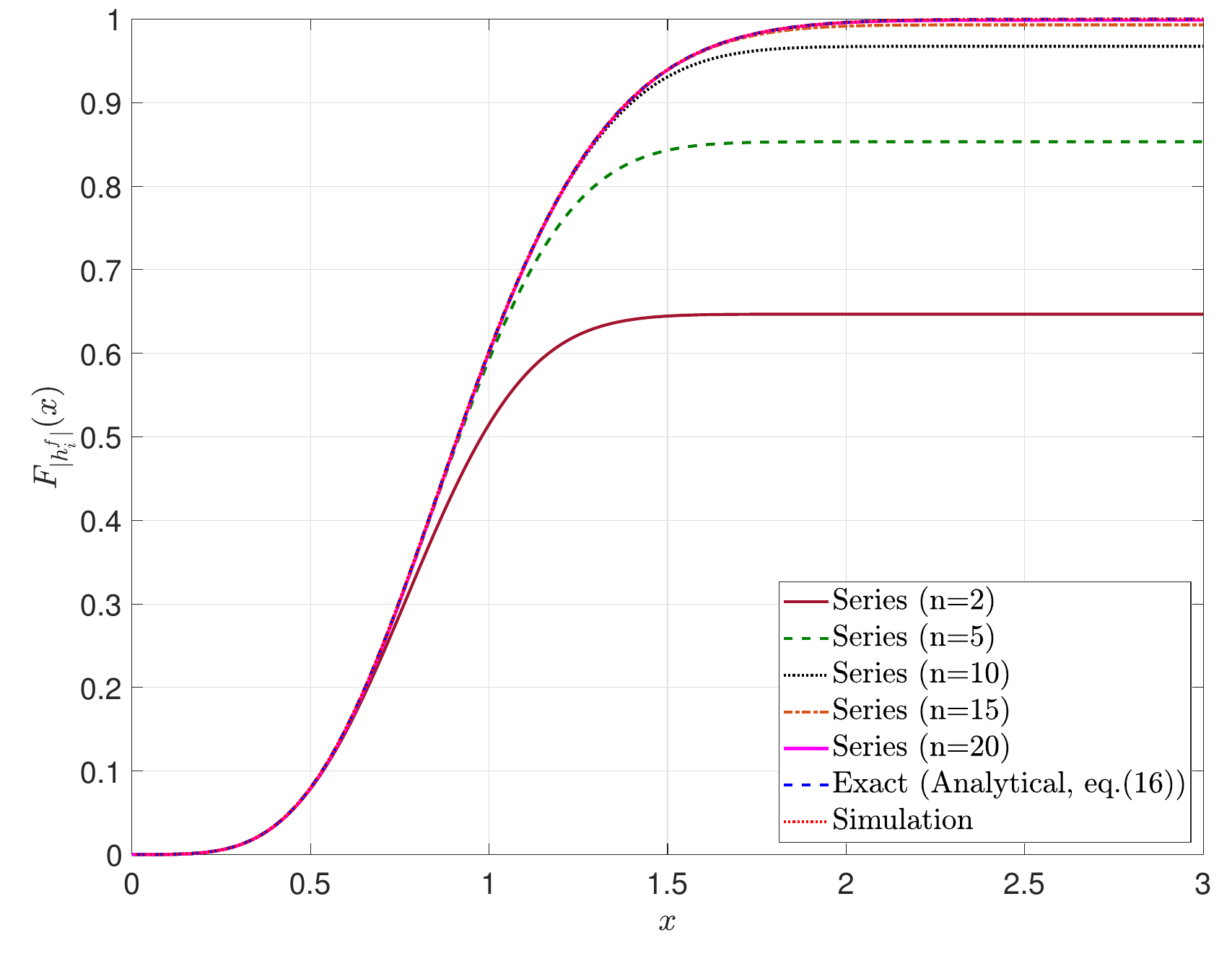}} 
		\caption{Density and distribution functions of the envelope for  $\alpha$-$\eta$-$\kappa$-$\mu$ fading channel: comparison between the derived exact  and infinite-series \cite{Silva_2020_alpha_eta_kappa_mu}.}
		\label{fig:pdf_cdf}	
	\end{figure*}
	
		\begin{figure*}[t]
		\centering
		\subfigure[Outage Probability, $p=1$, $q=1$, $\eta=1.01$]{\includegraphics[scale=0.3]{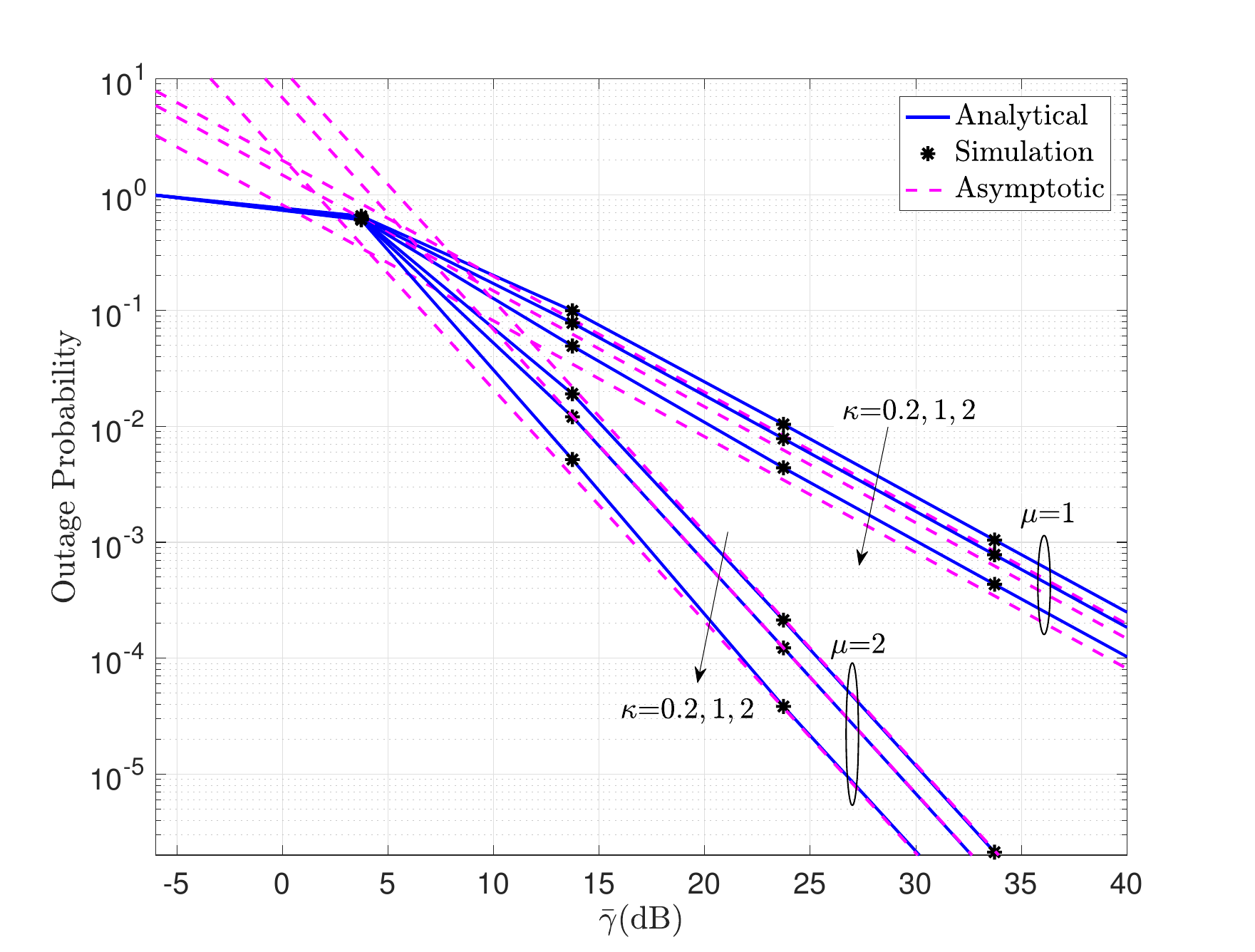}} 
		\subfigure[Average BER, $p=3$, $q=1$, $\mu=2$, $\kappa=1$]{\includegraphics[scale=0.3]{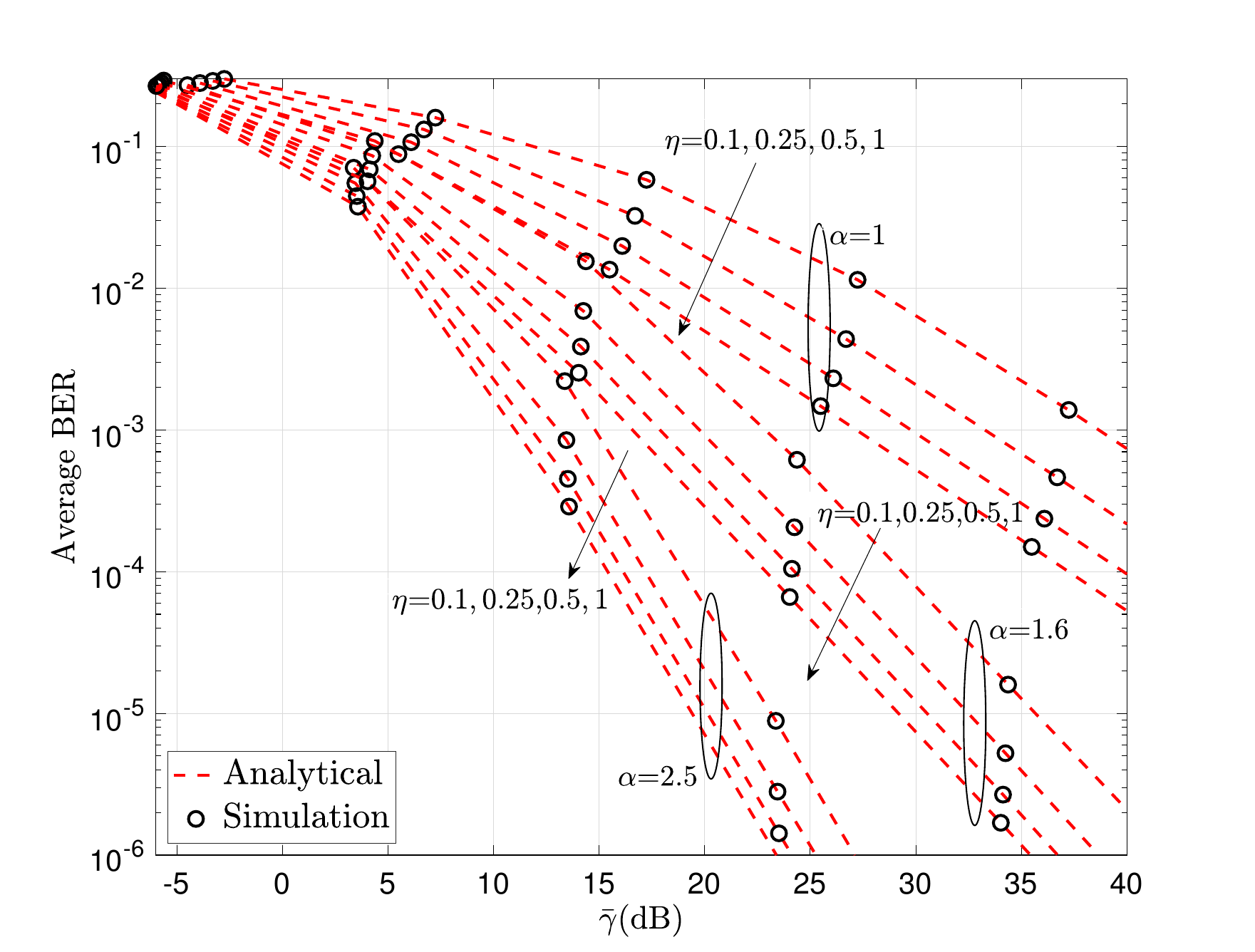}} 
		\caption{Outage probability and average BER performance of the single element ($N=1$) RRS system.}
		\label{fig:outage_ber}	
	\end{figure*}

	\section{Simulation and Numerical Results}
In this section, we use computer simulations over the generalized fading model to evaluate the developed statistical results for single-element and multiple-element RRS systems.  We perform numerical and simulation analysis to demonstrate the superiority of these results over the existing infinite series representation of the $\alpha$-$\eta$-$\kappa$-$\mu$ channel model.  We validate the derived analytical results using Monte Carlo simulations, employing the Fox's H-function code from \cite{Alhennawi2016}. To generate random samples of the envelope  $h_i^f| $ in \eqref{eq:Ralpha}, characterized by $\alpha, \eta, \kappa, \mu, p, q, \text{ and } \hat{r}$,  
we utilize the parametric equations from \cite{Yacoub_2016_alpha_eta_kappa_mu} provided in \eqref{eq:parameter}.

First, we analyze the single-element RRS system to demonstrate the effectiveness of the exact representation of the $\alpha$-$\eta$-$\kappa$-$\mu$ channel model, highlighting its accuracy and practical relevance. Next, we extend our study to the multi-element RRS system, examining its performance and the impact of multiple reflective elements on the system's overall behavior.
	\subsection{Simulation Results for Single-Element RRS}
In this subsection, we validate the exact representation of the PDF and CDF of the $\alpha$-$\eta$-$\kappa$-$\mu$ channel model and then apply it to evaluate outage probability and average BER under different channel conditions.
	
	In Fig. \ref{fig:pdf_cdf}, we numerically evaluate the derived PDF and CDF of the channel envelope using the software package of multivariate Fox's H-function \cite{Alhennawi2016}. We set $\alpha=2$, $\eta=1$, $\mu=2$, $\kappa=1$, $p=3$, $q=1$, and $\hat{r}=1$. The derived statistical results are validated through Monte Carlo simulations, where random samples of the envelope $R$ are generated, and the MATLAB function \emph{histogram ([$\zeta$], 'Normalization', 'pdf')} is used, with $[\zeta]$ representing an array of $10^7$ samples of the envelope $h_i^f| $. Further, we compare the derived analytical results with the state-of-the-art by numerically evaluating the infinite series PDF in \cite[eq. 11]{Silva_2020_alpha_eta_kappa_mu} (as given in \eqref{eq:pdf_series}) and the infinite series CDF in \cite[eq. 12]{Silva_2020_alpha_eta_kappa_mu} for different numbers of summands in \eqref{eq:pdf_series}. The statistical parameters for $h_i^f|$ in \eqref{eq:Ralpha} are determined by $\alpha, \eta, \kappa, \mu, p, q, \text{ and } \hat{r}$:	
	\begin{flalign} \label{eq:parameter}
	&\sigma_x = \sqrt{\frac{\eta (p+1)\hat{r}^\alpha}{2(\eta+1)(\kappa+1)\mu p}}, ~~ \sigma_y = \sqrt{\frac{ (p+1)\hat{r}^\alpha}{2(\eta+1)(\kappa+1)\mu p}}\nonumber \\ & 	\lambda_x = \sqrt{\frac{\eta \kappa q \hat{r}^\alpha}{(q\eta+1)(\kappa+1)}}, ~~~~~\lambda_y = \sqrt{\frac{ \kappa \hat{r}^\alpha}{(q\eta+1)(\kappa+1)}} \nonumber \\ &  \mu_x = \frac{2p\mu}{1+p}, ~~~~~~~~~~~~~~~~~~~~~  \mu_y = \frac{2\mu}{1+p}. 
	\end{flalign}
	\normalsize
	
Fig. \ref{fig:pdf_cdf}(a) illustrates that the infinite-series PDF significantly underestimates the actual PDF when the number of summands is small (e.g., $n=2$ and $n=5$) for the given system parameters. Even for a higher number of summands, such as $n=15$, discrepancies remain, particularly in the tail region of the distribution function for $x \geq 1.5$. While the infinite-series representation is ultimately convergent, achieving near-exact results within $20$ summands, the required number of summands for accurate convergence varies with channel parameters. This variation implies that an insufficient number of summands may lead to an inaccurate approximation of the system performance under different environmental conditions. Furthermore, Fig. \ref{fig:pdf_cdf}(a) demonstrates that the asymptotic PDF derived in \eqref{eq:asympt_pdf} closely follows the exact PDF when $x \to 0$, similar to the asymptotic PDF in \cite[eq.19]{Silva_2020_alpha_eta_kappa_mu}. However, unlike the asymptotic PDF in \cite[eq.19]{Silva_2020_alpha_eta_kappa_mu}, the derived asymptotic PDF also captures the overall shape of the actual PDF more accurately. In Fig. \ref{fig:pdf_cdf}(b), we further validate the accuracy of the derived CDF in \eqref{eq:outage_new_without_pe} by comparing it with the infinite-series CDF of \cite{Silva_2020_alpha_eta_kappa_mu}. The results exhibit similar trends and conclusions as observed in Fig. \ref{fig:pdf_cdf}(a), reinforcing the significance of the proposed exact and asymptotic representations.

	\begin{figure}	
		{\includegraphics[scale=0.3]{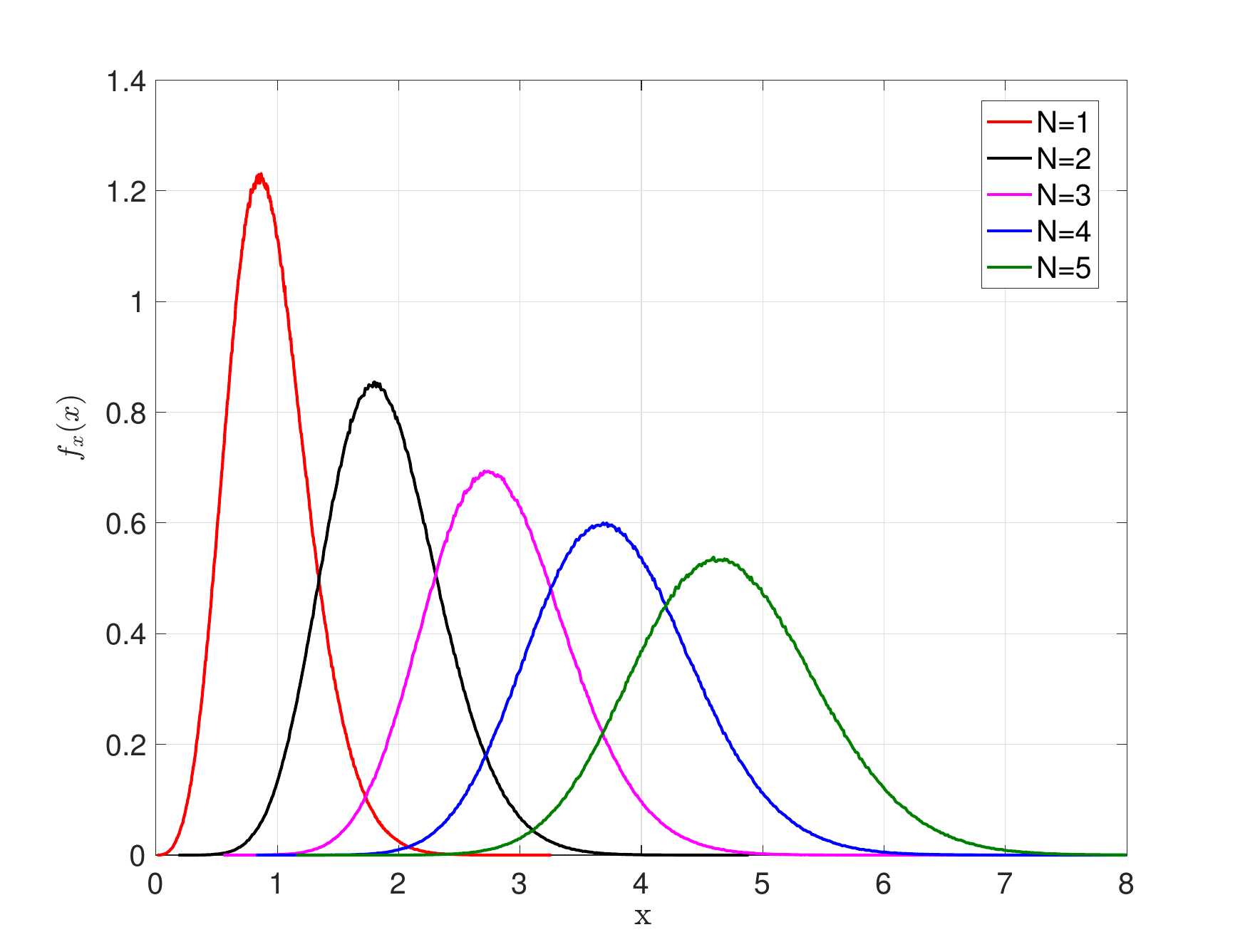}}. 
		\caption{PDF of RRS system  $Z=\sum_{i=}^N Z_i$  having parameters $\alpha=2$, $\eta=1$, $\kappa=1$, and $\mu=2$.}	
		\label{pdf_mrc}
	\end{figure}

	\begin{figure*}[t]
		\centering
		\subfigure[Outage Probability for different $N$]{\includegraphics[scale=0.3]{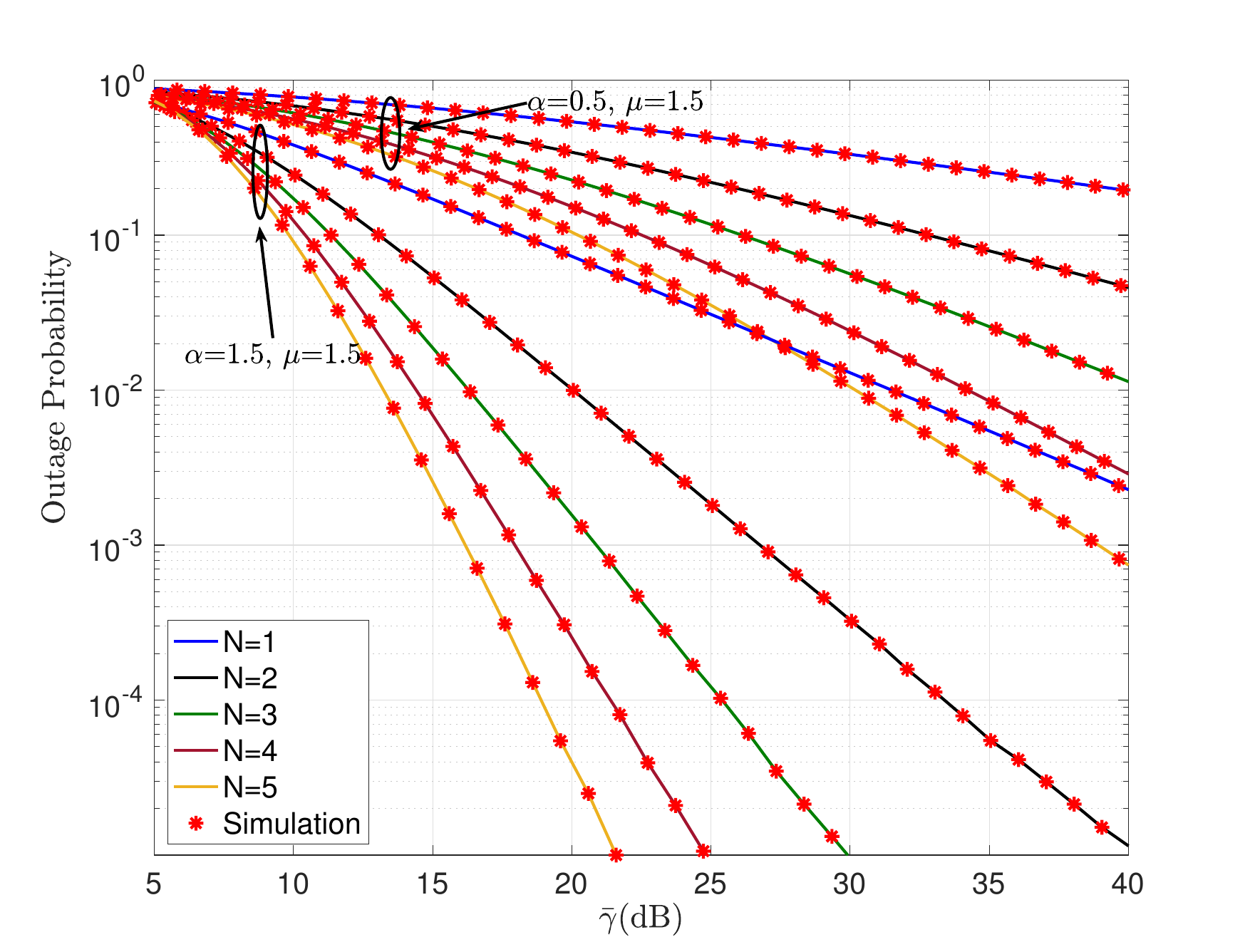}} 
		\subfigure[Average BER for different $N$]{\includegraphics[scale=0.3]{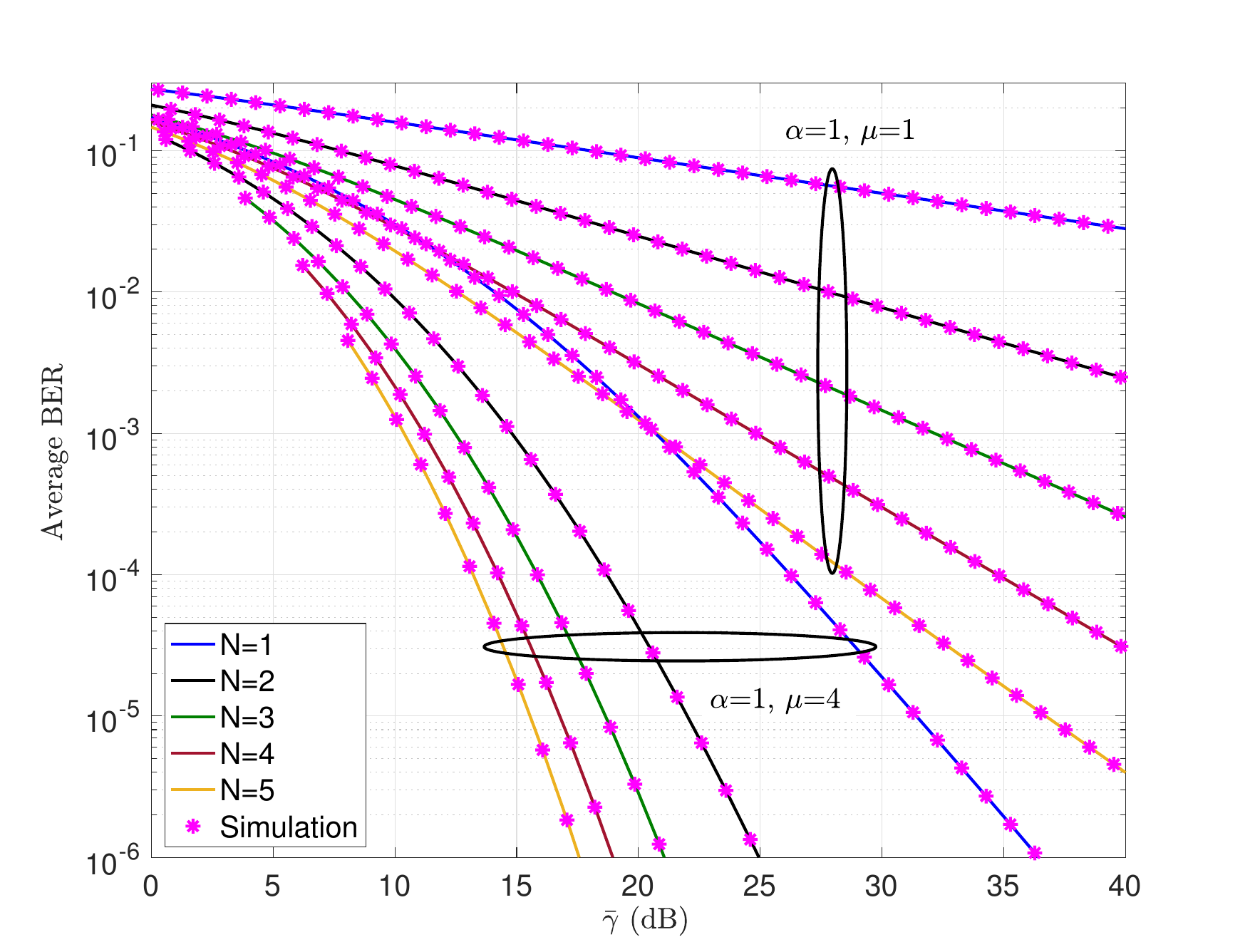}} 
		\caption{Outage probability and average BER performance of the RRS system versus   average SNR at different $N$  with parameters $p=3$, $q=1$, $\eta=1$, and $\kappa=1$.}
		\label{fig:outage_ber_mrc}	
	\end{figure*}

		\begin{figure*}[t]
		\centering
		\subfigure[Outage Probability ]{\includegraphics[scale=0.30]{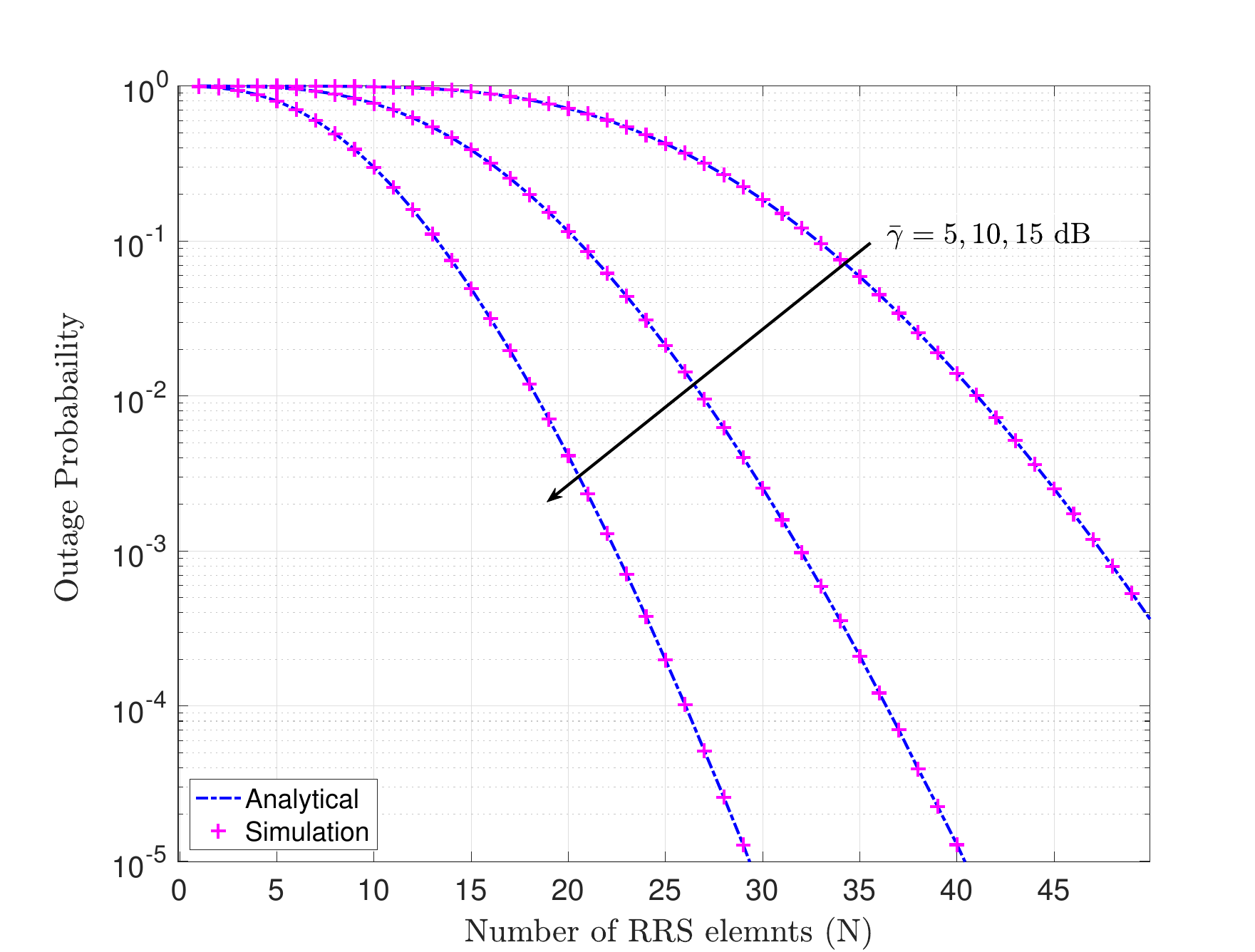}} 
		\subfigure[Average BER ]{\includegraphics[scale=0.30]{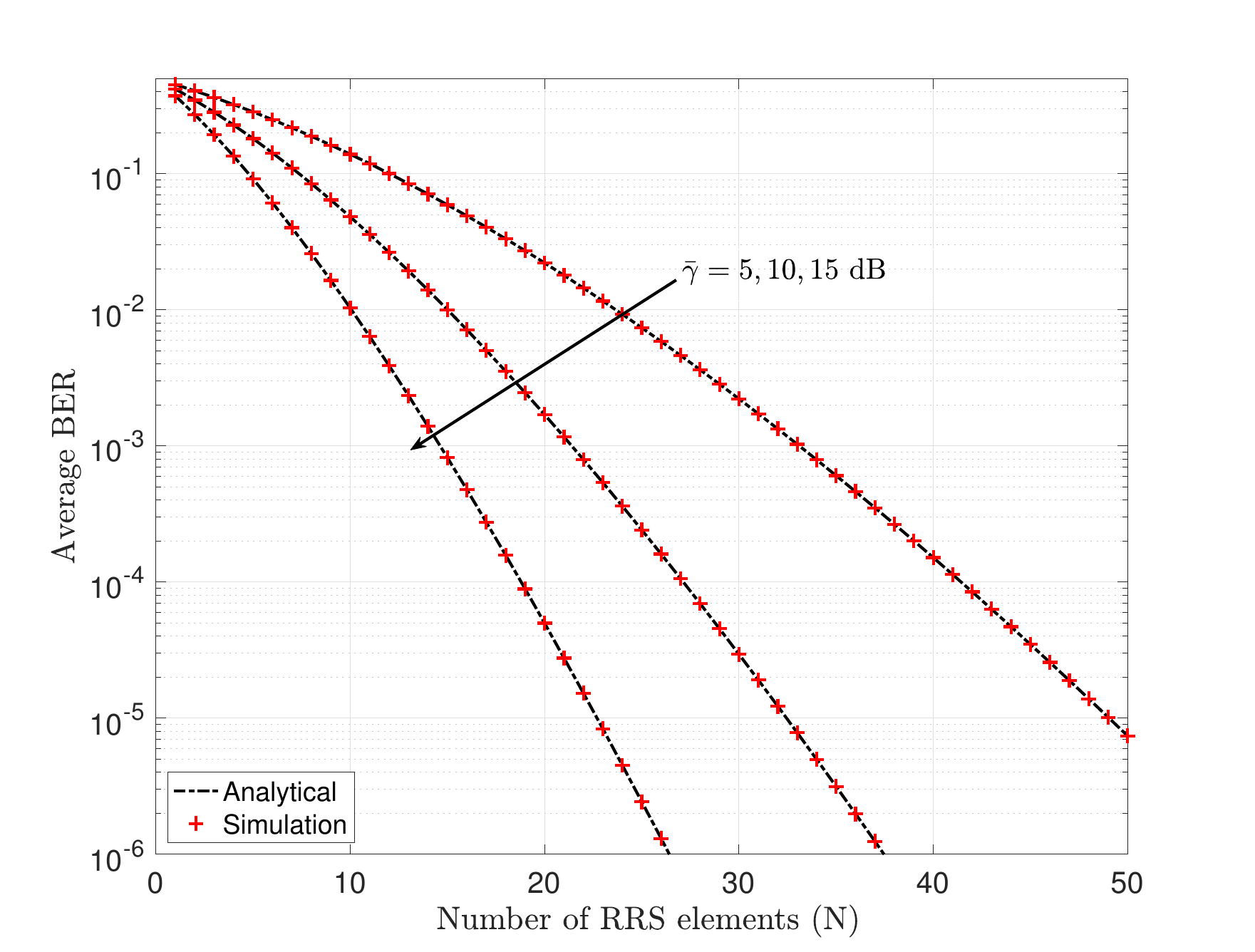}} 
		\caption{Outage probability and average BER performance of the RRS system versus $N$ at different average SNR with parameters  $\alpha=1$, $\mu=0.7$, $p=3$, $q=1$, $\eta=0.1$, and $\kappa=0.2$.}
		\label{fig:outage_ber_rrs_vs_N}	
	\end{figure*}
	
Understanding the outage and average BER performance is crucial for evaluating the reliability and efficiency of a single-element RRS system under various fading conditions. In Fig. \ref{fig:outage_ber}, we demonstrate the outage and average BER performance of a single-element RRS system, providing insights into its behavior across different channel parameters and system configurations.
Fig. \ref{fig:outage_ber}(a) depicts the outage performance of the system for different values of clustering parameter $\mu$ and the parameter $\kappa$. The outage performance of the system improves with an increase in the parameter $\mu$, signifying an increase in the number of multi-path clusters. The outage probability also decreases with an increase in $ \kappa $ due to the improvement in the power of the dominant component compared with the scattered waves. Fig. \ref{fig:outage_ber}(a) shows that the outage probability decreases from  $2.6\times 10^{-3}$ to $1.3\times10^{-5}$ when  $ \mu $ is increased from $1$ to $2$ with $ \kappa=0.2 $ at $30$ \mbox{dB} of average SNR. However, the improvement in the outage probability with an increase in the parameter $\kappa$ is not significant: the outage probability decreases from  $1.3\times 10^{-5}$ to $1.5\times10^{-6}$ when $\kappa$ is increased from $0.2$ to $2$ with $\mu=2$  at $30$ \mbox{dB} average SNR. Further, the derived asymptotic results for outage probability match very closely with analytical and simulated results at a reasonably high SNR. It should be emphasized that the slope of the outage probability changes with $\mu$ and remains constant with $\kappa$ verifying the diversity order (i.e., $\frac{\alpha\mu}{2}$) for the $\alpha$-$\eta$-$\kappa$-$\mu$ fading channel.

In Fig. \ref{fig:outage_ber} (b), we illustrate the average BER performance of the single-element RRS system by varying $\alpha$ and $\eta$ parameters for the BPSK modulation $\{ p_m  = 0.5, q_m  = 1\}$. The figure shows that the average BER decreases with an increase in $\alpha$ since the wireless channel tends to become more linear. Moreover, an increase in the value of $\eta$ reduces the average BER, as demonstrated in the existing literature. The figure shows that the average BER improves by $10$ when the value of $\eta$ increases from $0.1$ to $1$ with $\alpha=1$ at $30$ \mbox{dB} of average SNR. However, the non-linearity factor $\alpha$ changes the average BER more significantly: the average BER reduces from $6.5\times10^{-3}$ to $8\times10^{-5}$ (a factor $80$ reduction) when $\alpha =1$ (highly non-linear channel) increases to $\alpha =2.5$ (less non-linear channel) with $\eta=0.1$ at  $30$ \mbox{dB} of average SNR.

	\subsection{Simulation Results for Multiple-Element RRS} 
	In this subsection, we present the simulation results for the multiple-element RRS system to evaluate its performance in terms of PDF, outage probability, and average BER under various system parameters.
	Fig.~\ref{pdf_mrc}  illustrates the PDF of $Z$ for different values of $N$, with parameters $\alpha = 2$, $\eta = 1$, $\kappa = 1$, and $\mu = 2$. The results show that as $N$ increases, the distribution shifts towards higher values of $x$ and becomes more spread out, indicating an increase in the mean and variance of the received signal. This behavior suggests that adding more RRS elements enhances signal strength and diversity, leading to improved system performance by reducing the probability of deep fades and enhancing the overall reliability of the wireless link.
	
Fig. \ref{fig:outage_ber_mrc} presents the outage probability and average BER performance as functions of the average SNR for different numbers of RRS elements, ranging from $N=1$ to $N=5$, with parameters $p=3$, $q=1$, $\eta=1$, and $\kappa=1$.
Fig. \ref{fig:outage_ber_mrc}(a) illustrates the outage probability performance for different values of $\alpha$ and varying numbers of RIS elements $N$. The plot demonstrates the significant impact of increasing the number of RRS elements ($N$) on outage probability performance. For $\alpha = 1.5, \mu = 1.5$, achieving an outage probability of $10^{-4}$ requires an average SNR of approximately $19$\mbox{dB} when $N=5$, whereas for $N=1$, the required SNR exceeds $40$ \mbox{dB}, highlighting the substantial gain from diversity. Similarly, for an outage probability of $10^{-2}$, the required SNR is $11$\mbox{dB} for $N=3$ but increases to $20$ \mbox{dB} for $N=1$. When considering a more severe fading scenario ($\alpha = 0.5, \mu = 1.5$), the performance degrades significantly, as the system requires a much higher SNR to achieve the same outage probability. For example, at $10^{-3}$ outage probability, $N=5$ requires around $24$ \mbox{dB}, whereas $N=1$ needs more than $35$ \mbox{dB}, confirming that higher $\alpha$ values mitigate fading severity more effectively. The slope variation in the outage probability curves further confirms the theoretical diversity order, with steeper declines observed for larger $N$, indicating improved robustness against fading. The results collectively demonstrate that increasing $N$ significantly reduces the power requirements for reliable communication, while higher values of $\alpha$ and $\mu$ enhance system performance by suppressing fading effects.

The plot Fig. \ref{fig:outage_ber_mrc} (b)  illustrates the impact of increasing the number of RRS elements ($N$) on the average BER performance under different fading conditions. A key observation is that increasing $N$ significantly enhances signal quality, leading to a lower BER for a given SNR. For $\alpha = 1, \mu = 4$, achieving an average BER of $10^{-5}$ requires approximately $18$\mbox{dB} for $N=5$, whereas for $N=1$, the required SNR exceeds $30$\mbox{dB}. Similarly, for an average BER of $10^{-3}$, the required SNR is $12$\mbox{dB} for $N=3$, but it rises to $22$\mbox{dB} for $N=1$, emphasizing the role of RRS in improving system performance. The effect of the fading parameter $\mu$ is also evident; when comparing $\alpha = 1, \mu = 1$ and $\alpha = 1, \mu = 4$, it is observed that higher values of $\mu$ lead to a steeper BER decline. For instance, at an average BER of $10^{-4}$, the system with $N=5$ requires approximately $24$\mbox{dB} for $\mu=1$, whereas it only needs $17$\mbox{dB} for $\mu=4$. This confirms that larger $\mu$ values reduce the fading severity, improving BER performance. Further, the BER curves exhibit sharper slopes as $N$ increases, demonstrating that deploying more RRS elements provides enhanced signal strength and robustness against fading. These results highlight the importance of optimizing both $N$ and $\mu$ to achieve lower BER, minimize power consumption, and enhance communication reliability in fading environments.

Finally, we emphasize the impact of increasing $N$ on enhancing the performance of RRS-based systems, showcasing its ability to improve reliability and counteract fading-induced impairments. In Fig. \ref{fig:outage_ber_rrs_vs_N}, we present the outage probability and average BER of the RRS-assisted wireless system as a function of the number of RRS elements for $\alpha=1$, $\mu=0.7$, $p=3$, $q=1$, $\eta=0.1$, and $\kappa=0.2$. 

Fig. \ref{fig:outage_ber_rrs_vs_N} (a) demonstrates the effect of increasing the number of RRS elements ($N$) on outage probability for different SNR levels, $\bar{\gamma} = 5, 10, 15$\mbox{dB}. The results indicate that higher values of $N$ substantially enhance system reliability by reducing outage probability. Specifically, for $\bar{\gamma} = 15$\mbox{dB}, the outage probability decreases from approximately $10^{-1}$ at $N = 15$ to below $10^{-4}$ when $N$ surpasses $40$, signifying a drastic improvement. Similarly, for $\bar{\gamma} = 10$\mbox{dB}, an outage probability of $10^{-3}$ is attained at approximately $N = 30$, whereas for $\bar{\gamma} = 5$\mbox{dB}, even $N=50$ does not suffice to reach the same performance, indicating the need for additional RRS elements at lower SNR levels. The diminishing outage probability with increasing $N$ confirms that additional reflecting elements reinforce signal strength, effectively compensating for fading and improving wireless link reliability. The consistency between analytical and simulation results further supports the theoretical findings.

The plot in Fig. \ref{fig:outage_ber_rrs_vs_N} (b) depicts the variation of average BER with $N$ for different values of $\bar{\gamma}$. A clear trend is observed where a larger number of RRS elements leads to a notable reduction in BER, confirming the role of RRS in mitigating bit errors. Notably, the system's BER improves by approximately $80$ times when the average SNR increases from $0$\mbox{dB} to $5$\mbox{dB} for $N=30$, demonstrating the system's sensitivity to SNR variations. Furthermore, at $\bar{\gamma} = 15$\mbox{dB}, the BER decreases from nearly $10^{-2}$ at $N=10$ to below $10^{-5}$ when $N$ exceeds $40$, underscoring the substantial performance gain with additional RRS elements. Likewise, for $\bar{\gamma} = 10$\mbox{dB}, a BER of $10^{-4}$ is achieved around $N=35$, whereas for $\bar{\gamma} = 5$\mbox{dB}, even $N=50$ is insufficient to reach the same performance level. These results indicate that a combination of a higher SNR and an increased number of RRS elements is essential for maintaining low average BER performance. The observed improvement in the outage probability and BER performance with increasing $N$ confirms that leveraging a larger number of RRS elements effectively mitigates error rates, making it a promising approach for reliable communication in fading environments.

\section{Conclusion}
This paper presented an exact statistical analysis of RRS-based wireless transmission under the generalized $\alpha$-$\eta$-$\kappa$-$\mu$ fading model, considering both near-field transmission effects and far-field channel variations. We derived the exact PDF and CDF for the fading envelope using a single Fox-H function without infinite-series approximations, enabling an accurate performance assessment of wireless systems. The derived analytical expressions provided a framework for developing asymptotic outage probability analysis in the high SNR regime using simpler Gamma functions.  We utilized the statistical results of the single-channel model to analyze the $N$-element RRS-based transmission, expressing the sum of trivariate Fox-H functions in terms of a $3N$-variate Fox-H function. To validate the superiority of the proposed analysis, we derived the exact outage probability and average BER for both single-element and multiple-element RRS systems under $\alpha$-$\eta$-$\kappa$-$\mu$ fading conditions. Monte Carlo simulations confirmed the accuracy of the proposed statistical model, showing a close match with analytical results. The findings revealed that the conventional infinite-series representation of the $\alpha$-$\eta$-$\kappa$-$\mu$ fading model led to significant errors when using a limited number of summands, particularly in the tail region of the fading distribution.  Furthermore, increasing the number of RRS elements ($N$) significantly reduced outage probability and improved BER performance, confirming the diversity and power gains of RRS-assisted systems. For instance, at 15 dB SNR, the outage probability decreased from $10^{-1}$ ($N=15$) to below $10^{-4}$ ($N>40$), demonstrating enhanced robustness against fading impairments. Similarly, the BER performance improved significantly, with reductions of up to 80 times as the number of RRS elements increased. Further, the impact of key channel parameters such as $\mu, \kappa, \alpha$, and $\eta$ was analyzed, confirming that higher values of $\mu$ and $\alpha$ significantly enhanced reliability and BER performance by mitigating fading effects.  

We anticipate that the proposed statistical analysis of the fading envelope will revitalize interest in the $\alpha$-$\eta$-$\kappa$-$\mu$ model, given its excellent agreement with experimental data and improved mathematical tractability for performance evaluation in next-generation wireless systems. Future research could focus on experimental validation through real-world channel measurements to further substantiate the theoretical findings. Moreover, a comparative study between RRS and massive MIMO systems could be conducted to assess their respective benefits in beyond-5G and 6G wireless networks.

		\section{Acknowledgment}
	The authors would like to thank Ms. E.S. Karnawat,  Mr. Jaival Talati, and  Mr. Adwait Vyahalkar for their help in earlier versions of this paper.

	\section*{Appendix A}
We use the contour-integral representation of exponential function in $s_1$ and hyper-geometric function in $s_2$ to get the MGF of $Z_i$ as
	\begin{align}\label{eq:mgf_18}
		&M_{Z_i}(t) =  \left(\frac{1}{2\pi \J}\right)^2 \frac{\alpha_i (\xi_i \mu_i)^\mu_i}{{(g_i\hat{r}_i})^{\alpha_i\mu_i}\exp\Big(\frac{(1+p_iq_i)\kappa_i\mu_i}{\delta_i}\Big)} \bigg(\frac{p_i}{\eta_i}\bigg)^{\frac{p_i\mu_i}{1+p_i}} \nonumber \\& \times  \sum_{n=0}^{\infty}\Big(\frac{\xi_i \mu_i (p_i-\eta_i)}{(g_i\hat{r}_i)^{\alpha_i} \eta_i}\Big)^n L_n^{\frac{\mu_i}{1+p_i}-1} \Big(\frac{\eta_i \kappa_i \mu_i}{\delta_i (\eta_i-p_i)}\Big) \nonumber \\&\times \int_{\mathcal{L}_{i,1}} \int_{\mathcal{L}_{i,2}}\int_{0}^{\infty} e^{-tx} x^{{\alpha_i}({\mu_i+n+ s_1+\ s_2)-1}}  dx   \Gamma(-s_1)\nonumber \\&\times \Big(\frac{p_i \xi_i \mu_i}{(g_i\hat{r}_i)^{\alpha_i} \eta_i}\Big)^{s_1} \frac{ \Gamma(- s_2)}{ \Gamma(\mu_i + \eta_i + s_2)} (\frac{p_i^2q_ix^{\alpha_i} \kappa_i \xi_i \mu_i^2}{(g_i\hat{r}_i)^{\alpha_i} \delta_i \eta_i}\Big)^{s_2} d{s_2} d_{s_1}
	\end{align}
	Solving the inner integral $ \int_{0}^{\infty}e^{-tr}  r^{\alpha(\mu + n + s_1 + s_2)-1} dr $ \cite[3.381.4]{Gradshteyn}, we represent \eqref{eq:mgf_18} as
	\begin{align}\label{eq:mgf_19}
		&M_{Z_i}(t) = \left(\frac{1}{2\pi \J}\right)^2  \frac{\alpha_i (\xi_i \mu_i)^\mu_i}{{(g_i\hat{r}_i})^{\alpha_i\mu_i}\exp\Big(\frac{(1+p_iq_i)\kappa_i\mu_i}{\delta_i}\Big)} \bigg(\frac{p_i}{\eta_i}\bigg)^{\frac{p_i\mu_i}{1+p_i}}  \nonumber \\&\times \sum_{n=0}^{\infty}\frac{1}{t^{\alpha_i n}}\Big(\frac{\xi_i \mu_i (p_i-\eta_i)}{(g_i\hat{r}_i)^{\alpha_i} \eta_i}\Big)^n L_n^{\frac{\mu_i}{1+p_i}-1} \Big(\frac{\eta_i \kappa_i \mu_i}{\delta_i (\eta_i-p_i)}\Big) \nonumber \\& \times\int_{\mathcal{L}_{i,1}} \int_{\mathcal{L}_{i,2}}\frac{\Gamma(-s_1)\Gamma(-s_2)\Gamma{(\alpha_i(\mu_i + n + s_1 + s_2))}}{\Gamma(\mu_i + \eta_i + s_2)}  \nonumber \\&\times\Big(\frac{p_i \xi_i \mu_i}{(t g_i\hat{r}_i)^{\alpha_i} \eta_i}\Big)^{s_1}(\frac{p_i^2q_ix^{\alpha_i} \kappa_i \xi_i \mu_i^2}{(t g_i\hat{r}_i)^{\alpha_i} \delta_i \eta_i}\Big)^{s_2} d{s_2} d_{s_1}
	\end{align}
	Using \eqref{eq:mgf_19} in	\eqref{eqn:pdf_mgf_main},  the PDF of $Z$ can be expressed as
	\begin{flalign}\label{eq:pdf_inf1}
		&	f_{Z}(x) = \left(\frac{1}{2\pi \J}\right)^{2N+1}  \sum_{n_1=0}^{\infty} \sum_{n_2=0}^{\infty} \sum_{n_3=0}^{\infty} \cdots \sum_{n_N=0}^{\infty} \nonumber \\&\times\prod_{j=1}^{N} \frac{\alpha_j (\xi_j \mu_j)^{\mu_j}}{(g_j \hat{r}_j)^{\alpha_j \mu_j} \exp\left( \frac{(1 + p_j q_j) \kappa_j \mu_j}{\delta_j} \right)}\nonumber \\& \times\left( \frac{p_j}{\eta_j} \right)^{\frac{p_j \mu_j}{1 + p_j}} \left( \frac{\xi_j \mu_j (p_j - \eta_j)}{(g_j \hat{r}_j)^{\alpha_j} \eta_j} \right)^{n_j}
		L_n^{\frac{\mu_j}{1 + p_j} - 1} \left( \frac{\eta_j \kappa_j \mu_j}{\delta_j (\eta_j - p_j)} \right)\nonumber \\& \times	\int_{\mathcal{L}_{1,1}} \int_{\mathcal{L}_{1,2}} \cdots \int_{\mathcal{L}_{N,1}} \int_{\mathcal{L}_{N,2}}  \int_{\mathcal{L}} e^{tx}  \frac{1}{t^{\sum_{i=1}^N(\alpha_i n_i + \alpha_i s_{i,1} + \alpha_i s_{i,2})}} dt\nonumber \\& \times \prod_{i=1}^{N}\frac{\Gamma(-s_{i,1}) \Gamma(-s_{i,2}) \Gamma\left(\alpha_i(\mu_i + n_i + s_{i,1} + s_{i,2})\right)}{\Gamma(\mu_i + \eta_i + s_{i,2})} \nonumber \\&\times
		\left( \frac{p_i \xi_i \mu_i}{( g_i \hat{r}_i)^{\alpha_i} \eta_i} \right)^{s_{i,1}}
		\left( \frac{p_i^2 q_i x^{\alpha_i} \kappa_i \xi_i \mu_i^2}{( g_i \hat{r}_i)^{\alpha_i} \delta_i \eta_i} \right)^{s_{i,2}}
		d{s_{i,2}} d{s_{i,1}}
	\end{flalign}
	Solving the inner integral by applying the identity [], we simplify \eqref{eq:pdf_inf1}:
	\begin{flalign}
		&	f_{Z}(x) = \sum_{n_1=0}^{\infty} \sum_{n_2=0}^{\infty} \sum_{n_3=0}^{\infty} \cdots \sum_{n_N=0}^{\infty} \nonumber \\&\times x ^{ \sum_{i=1}^N(\alpha_i n_i -1)} \prod_{j=1}^{N} \frac{\alpha_j (\xi_j \mu_j)^{\mu_j}}{(g_j \hat{r}_j)^{\alpha_j \mu_j} \exp\left( \frac{(1 + p_j q_j) \kappa_j \mu_j}{\delta_j} \right)}\nonumber \\&\times \left( \frac{p_j}{\eta_j} \right)^{\frac{p_j \mu_j}{1 + p_j}} \left( \frac{\xi_j \mu_j (p_j - \eta_j)}{(g_j \hat{r}_j)^{\alpha_j} \eta_j} \right)^n
		L_n^{\frac{\mu_j}{1 + p_j} - 1} \left( \frac{\eta_j \kappa_j \mu_j}{\delta_j (\eta_j - p_j)} \right) \nonumber \\& \times	\left(\frac{1}{2\pi \J}\right)^{2N} \int_{\mathcal{L}_{1,1}} \int_{\mathcal{L}_{1,2}} \cdots \int_{\mathcal{L}_{N,1}} \int_{\mathcal{L}_{N,2}} \nonumber \\& \times \frac {x ^{ \sum_{i=1}^N( \alpha_i s_{i,1} + \alpha_i s_{i,2})}} {{\Gamma\left( \sum_{i=1}^N(\alpha_i n + \alpha_i s_{i,1} + \alpha_i s_{i,2})\right)}} \nonumber \\& \times \prod_{i=1}^{N}\frac{\Gamma(-s_{i,1}) \Gamma(-s_{i,2}) \Gamma\left(\alpha_i(\mu_i + n_i + s_{i,1} + s_{i,2})\right)}{\Gamma(\mu_i + \eta_i + s_{i,2})} \nonumber \\& \times
		\left( \frac{p_i \xi_i \mu_i}{( g_i \hat{r}_i)^{\alpha_i} \eta_i} \right)^{s_{i,1}}
		\left( \frac{p_i^2 q_i x^{\alpha_i} \kappa_i \xi_i \mu_i^2}{( g_i \hat{r}_i)^{\alpha_i} \delta_i \eta_i} \right)^{s_{i,2}}
		d{s_{i,2}} d{s_{i,1}}
	\end{flalign}
	Applying the definition of multivariate Fox's H function \cite{Mathai_2010} we get the PDF in \eqref{th:mgf_infinity} to conclude the proof of Theorem \ref{th:mgf_infinity}.
	\section*{Appendix B}
	Using $ f_U (u) $ and $ f_V (v) $ from \cite{Silva_2020_alpha_eta_kappa_mu} in \eqref{eq:derivation_1} and after some algebraic 
	manipulation, we get\footnote{There is a typo in equation (9) of \cite{Silva_2020_alpha_eta_kappa_mu}. It should be $(\hat{r}^\alpha)^{1+\frac{\mu}{2}}$ in the denominator of the second term of the first row.} \small
	\begin{flalign} \label{eq:pdf_eq9}
	f_{|h_i^f|}(x) &= \frac{\psi_1 x^{\psi_2}}{(\hat{r}^\alpha)^{1+\frac{\mu}{2}}} \times e^{-\psi_3 x^{\alpha}} \int_{0}^{x^\alpha}
	\frac{(x^{\alpha}-v)^{\frac{A_1}{2}}}{v^{-\frac{A_2}{2}}}  e^{-A_{3}v} \nonumber \\ & \times
	I_{A_{1}}(A_{4}(x^\alpha-v)^{\frac{1}{2}}) I_{A_{2}}(A_{5}v^{\frac{1}{2}})dv
	\end{flalign} \normalsize
	where $I_{A_{1}}$ and $I_{A_{2}}$ are the modified Bessel function of the first kind. Using Meijer's G representation of the Bessel functions \cite{Mathematica_besseli}, \eqref{eq:pdf_eq9} can be rewritten as
	\small
	\begin{flalign} \label{eq:pdf_derive_1}
	f_{|h_i^f|}(x)&=  \frac{\psi_1 x^{\psi_2}}{(\hat{r}^\alpha)^{1+\frac{\mu}{2}}}  e^{-\psi_3x^\alpha} \pi 2^{-A_1} (A_{4}(x^\alpha-v)^{\frac{1}{2}})^{A_1} \nonumber\\ & \times  \pi 2^{-A_2} (A_{5}v^{\frac{1}{2}})^{A_2}  \int_{0}^{x^\alpha}\frac{(x^{\alpha}-v)^{\frac{A_1}{2}}}{v^{-\frac{A_2}{2}}}  G_{0,1}^{1,0}\left(\begin{array}{c} -\\0 \end{array}\left| A_3 v\right. \right) \nonumber \\ & \times G_{1,3}^{1,0} \left(\begin{array}{c}\frac{1}{2}\\0, -A_1, \frac{1}{2} \end{array} \left | \frac{A_{4}^2(x^\alpha -v)}{4}\right. \right)  \nonumber\\ & \times   G_{1,3}^{1,0} \left (\begin{array}{c} \frac{1}{2}\\0, -A_2, \frac{1}{2} \end{array} \left| \frac{A_{5}^2v}{4}\right. \right) dv
	\end{flalign}
	\normalsize
	Utilizing the integral representation of Meijer's G-function \cite{Mathai_2010}, we can represent \eqref{eq:pdf_derive_1} as 
	
	\small
	\begin{flalign} \label{eq:pdf_derive_3}
	&f_{|h_i^f|}(x) = \frac{\psi_1}{(\hat{r}^\alpha)^{1+\frac{\mu}{2}}} \pi^2 2^{(2-\mu)} A_4^{A_1} A_5^{A_2} x^{\psi_2} e^{-\psi_3x^\alpha} \nonumber \\ & \times \frac{1}{(2\pi i)^{3}} \int_{\mathcal{L}_1}\int_{\mathcal{L}_2}\int_{\mathcal{L}_3}\Gamma(-s_1){A_3}^{s_1} ds_1 \nonumber \\ & \times \frac{\Gamma(-s_2)}{\Gamma(1+A_1+s_2)\Gamma(\frac{1}{2}+s_2)\Gamma(\frac{1}{2}-s_2)}\Big(\frac{A_{4}^{2}}{4} \Big)^{s_2} ds_2 \nonumber
	\\& \times \frac{\Gamma(-s_3)}{\Gamma(1+A_2+s_3)\Gamma(\frac{1}{2}+s_3)\Gamma(\frac{1}{2}-s_3)}\Big(\frac{A_{5}^{2}}{4} \Big)^{s_3} I_1 ds_3  
	\end{flalign}
	\normalsize
	where $\mathcal{L}_1$, $\mathcal{L}_2$, and $\mathcal{L}_3$, denote the contour integrals. The inner integral $I_1 $ can be represented and simplified using the identity \cite[3.191.1]{Gradshteyn} as
	\begin{flalign} \label{inn}
	&I_1 = \int_{0}^{r^\alpha} {v}^{{A_2}+s_1+s_3} (x^{\alpha}-v)^{{A_1}+s_2} dv  = \nonumber \\  & \frac{\Gamma(1+A_2+s_1+s_3)\Gamma(1+A_1+s_2)}{\Gamma(2+A_2+s_1+s_3+A_1+s_2)}  
	{x^{\alpha(1+A_2+s_1+s_3+A_1+s_2)}}
	\end{flalign}
	
	Finally, substituting  \eqref{inn} in \eqref{eq:pdf_derive_3}, rearranging the terms, and applying the definition of multivariate Fox's H-function \cite{Mathai_2010}, we get the PDF of Theorem 1 in \eqref{eq:pdf_new_without_pe}, which  concludes the proof. 
	
	\section*{Appendix C}

	The MGF of $Z_i$ can be obtained by standard transformation of random variables  \cite{papoulis_2002}:
	\small
	
	\begin{flalign} \label{eq:MGF_SNR_NEW}
		&M_{Z_i}(t) = |g_i|^{-\alpha\mu_i+1} \frac{\psi_1}{(\hat{r}^{\alpha_i})^{1+\frac{\mu_i}{2}}} \pi^2 2^{(2-\mu_i)} A_{i,4}^{A_{i,1}} A_{i,5}^{A_{i,2}} \nonumber \\ & \times \frac{1}{(2\pi \J)^{3}}\int_{\mathcal{L}_{i,1}}\int_{\mathcal{L}_{i,2}}\int_{\mathcal{L}_{i,3}} \int_{0}^{\infty}  e^{-tx}  x^{\alpha_i\mu_i-1+\alpha_i(s_{i,1}+s_{i,2}+s_{i,3})} \nonumber \\& \times e^{-\psi_3|g_i|^{-\alpha_i}x^{\alpha_i}} dx |g_i|^{-\alpha_i(s_{i,1}+s_{i,2}+s_{i,3})}\Gamma(-s_{i,1}){A_{i,3}}^{s_{i,1}}  \nonumber \\ & \times \frac{\Gamma(-s_{i,2})}{\Gamma(1+A_{i,1}+s_{i,2})\Gamma(\frac{1}{2}+s_{i,2})\Gamma(\frac{1}{2}-s_{i,2})}\Big(\frac{A_{i,4}^{2}}{4} \Big)^{s_{i,2}}  \nonumber
		\\& \times \frac{\Gamma(-s_{i,3})}{\Gamma(1+A_{i,2}+s_{i,3})\Gamma(\frac{1}{2}+s_{i,3})\Gamma(\frac{1}{2}-s_{i,3})}\nonumber \\& \times \Big(\frac{A_{i,5}^{2}}{4} \Big)^{s_{i,3}} \frac{\Gamma(1+A_{i,2}+s_{i,1}+s_{i,3})\Gamma(1+A_{i,1}+s_{i,2})}{\Gamma(2+A_{i,2}+s_{i,1}+s_{i,3}+A_{i,1}+s_{i,2})}  
		ds_{i,1} ds_{i,2}ds_{i,3}
	\end{flalign}	
	Applying the Meijer-G equivalent of the exponential function over contour integral $\mathcal{L}_{i,4}$ and using the definition of the Gamma function to solve the inner integral in variable $x$, we get
	\begin{flalign} \label{eq:MGF_SNR_NEW}
		&M_{Z_i}(t) = |g_i|^{-\alpha\mu_i+1} \frac{\psi_1}{(\hat{r}^{\alpha_i})^{1+\frac{\mu_i}{2}}} \pi^2 2^{(2-\mu_i)} A_{i,4}^{A_{i,1}} A_{i,5}^{A_{i,2}} \frac{1}{(2\pi \J)^{3}}\nonumber \\ & \times \frac{1}{(2\pi \J)^{3}}\int_{\mathcal{L}_{i,1}}\int_{\mathcal{L}_{i,2}}\int_{\mathcal{L}_{i,3}} \int_{\mathcal{L}_{i,4}}   t^{-\alpha_i (\mu_i+ s_{i,1} + s_{i,2} + s_{i,3} + s_{i,4})}\nonumber \\ & \times \Gamma(\alpha_i ( \mu_i +  s_{i,1} + s_{i,2} + s_{i,3} + s_{i,4}))) |g_i|^{-\alpha_i(s_{i,1}+s_{i,2}+s_{i,3})}  \nonumber \\ & \times\Gamma(-s_{i,1}){A_{i,3}}^{s_{i,1}} \frac{\Gamma(-s_{i,2})}{\Gamma(1+A_{i,1}+s_{i,2})\Gamma(\frac{1}{2}+s_{i,2})\Gamma(\frac{1}{2}-s_{i,2})}\nonumber
		\\& \times\Big(\frac{A_{i,4}^{2}}{4} \Big)^{s_{i,2}}   \frac{\Gamma(-s_{i,3})}{\Gamma(1+A_{i,2}+s_{i,3})\Gamma(\frac{1}{2}+s_{i,3})\Gamma(\frac{1}{2}-s_{i,3})}\Big(\frac{A_{i,5}^{2}}{4} \Big)^{s_{i,3}}\nonumber \\ & \times \frac{\Gamma(1+A_{i,2}+s_{i,1}+s_{i,3})\Gamma(1+A_{i,1}+s_{i,2})}{\Gamma(2+A_{i,2}+s_{i,1}+s_{i,3}+A_{i,1}+s_{i,2})}\nonumber \\& \times \Gamma(-s_{i,4}) \big(\psi_3|g_i|^{-\alpha_i}\big)^{s_{i,4}}  
		ds_{i,1} ds_{i,2}ds_{i,3}  ds_{i,4} 
	\end{flalign}	
	To get the PDF of the resultant channel for RRS-assisted transmission, 	we use \eqref{eq:MGF_SNR_NEW} in	\eqref{eqn:pdf_mgf_main}. Thus, we use the identity $\frac{1}{\Gamma(z)}=\frac{j}{2\pi}\int_{c}(-y)^{-z}e^{-y}dy$ for the inner integral in $t$ to get
	\begin{flalign}\label{eq:pdf_mv2}
		&f_{Z}(x)=\prod_{j=1}^{N}|g_j|^{-\alpha\mu_j+1} \frac{\psi_1}{(\hat{r}^{\alpha_j})^{1+\frac{\mu_j}{2}}} \pi^2 2^{(2-\mu_j)} A_{j,4}^{A_{j,1}} A_{j,5}^{A_{j,2}} \nonumber \\ &\times\frac{1}{(2\pi \J)^4} \int_{\mathcal{L}_{i,1}}\int_{\mathcal{L}_{i,2}}\int_{\mathcal{L}_{i,3}} \int_{\mathcal{L}_{i,4}} \cdots \int_{\mathcal{L}_{N,1}}\int_{\mathcal{L}_{N,2}}\int_{\mathcal{L}_{N,3}} \int_{\mathcal{L}_{N,4}} \nonumber \\& \times \frac{x^{\sum_{i=1}^{N}{\alpha_i ( \mu_i +  s_{i,1} + s_{i,2} + s_{i,3} + s_{i,4})}}}{\Gamma(\sum_{i=1}^{N}\alpha_i ( \mu_i +  s_{i,1} + s_{i,2} + s_{i,3} + s_{i,4}))} \nonumber \\& \times\prod_{i=1}^{N}\Gamma(\alpha_i ( \mu_i +  s_{i,1} + s_{i,2} + s_{i,3} + s_{i,4})) \Gamma(-s_{i,1})({A_{i,3}|g_i|^{-\alpha_i}})^{s_{i,1}} \nonumber \\ &  \times \frac{\Gamma(-s_{i,2})}{\Gamma(1+A_{i,1}+s_{i,2})\Gamma(\frac{1}{2}+s_{i,2})\Gamma(\frac{1}{2}-s_{i,2})}\Big(\frac{A_{i,4}^{2}|g_i|^{-\alpha_i}}{4} \Big)^{s_{i,2}}  \nonumber
		\\& \times \frac{\Gamma(-s_{i,3})}{\Gamma(1+A_{i,2}+s_{i,3})\Gamma(\frac{1}{2}+s_{i,3})\Gamma(\frac{1}{2}-s_{i,3})}\Big(\frac{A_{i,5}^{2}|g_i|^{-\alpha_i}}{4} \Big)^{s_{i,3}} \nonumber\\&\times\frac{\Gamma(1+A_{i,2}+s_{i,1}+s_{i,3})\Gamma(1+A_{i,1}+s_{i,2})}{\Gamma(2+A_{i,1}+A_{i,2}+s_{i,1}+s_{i,2}+s_{i,3})} \Gamma(-s_{i,4}) \nonumber\\&\times \big(\psi_3|g_i|^{-\alpha_i}\big)^{s_{i,4}}  
		ds_{i,1} ds_{i,2}ds_{i,3}  ds_{i,4} 
	\end{flalign}
	
	Utilizing the definition of multivariate Fox's H-function in \eqref{eq:pdf_mv}, we get the PDF of Theorem \ref{th:pdf_mrc} in \eqref{eq:mrc_pdf_foxh}.
	
		\bibliographystyle{IEEEtran}
	\bibliography{alpha_eta_kappa_mu,Mybibfile}
\end{document}